\def\SumD{\sum_{t=1}^n}
\def\SumA{\sum\limits_{{}_{t: y_t\in(0,1)}}\!\!\!\!}
\def\Reta_betar{{\partial\eta_t\over\partial\beta_r}}
\def\Seta_betas{{\partial\eta_t\over\partial\beta_s}}
\def\Teta_betau{{\partial\eta_t\over\partial\beta_u}}
\def\G_Mu2{{{\partial^2}\over{\partial\mu_t^2}}}
\def\P1_Fi{{\partial^2\ell_t(\mu_t,\phi)\over\partial\mu_t\partial\phi }}
\def\stacksymbols #1#2#3#4{\def\theguybelow{#2}
   \def\verticalposition{\lower#3pt}
   \def\spacingwithinsymbol{\baselineskip0pt\lineskip#4pt}
   \mathrel{\mathpalette\intermediary#1}}
\def\intermediary#1#2{\verticalposition\vbox{\spacingwithinsymbol
   \everycr={}\tabskip0pt
   \halign{$\mathsurround0pt#1\hfil##\hfil$\crcr#2\crcr
           \theguybelow\crcr}}}

\def\Xtil{\stacksymbols{{\cal X}}{\sim}{5}{.7}}

\def\Wtil{\stacksymbols{{\mathbf{W}}}{\sim}{5}{.7}}
\def\ytil{\stacksymbols{{\mathbf{y}}}{\sim}{5}{.7}}
\def\Ttil{\stacksymbols{{\mathbf{T}}}{\sim}{5}{.7}}

\def\R{\item}

\def\1DL_mu1{{\partial\ell_t  (\mu_t,\phi)\over\partial\mu_t }}
\def\2DL_mu2{{\partial^2\ell_t(\mu_t,\phi)\over\partial\mu_t^2}}
\def\3DL_mu3{{\partial^3\ell_t(\mu_t,\phi)\over\partial\mu_t^3}}

\documentclass[letterpaper, 11pt]{article} 

\usepackage[english]{babel} 
\usepackage[latin1]{inputenc}     

\usepackage{amsmath,amsthm}
\usepackage{amssymb,latexsym}
\usepackage{graphics}
\usepackage{graphicx}

\usepackage{booktabs} 

\usepackage{url}

\numberwithin{equation}{section}

\newcommand{\resumen}[2]{%
\begin{abstract}#1\\\\\noindent\textit{\textbf{Keywords and Phrases}: }#2\end{abstract}}

\title{\Large\bf A general class of zero-or-one inflated beta regression models\bigskip}
\author{
Raydonal Ospina\\
Departamento de Estatística/CCEN\\ 
Universidade Federal de Pernambuco\\
Cidade Universitária, Recife/PE 50740-540, Brazil.\\
{\tt e-mail: raydonal@de.ufpe.br} \\ \\
Silvia L.P. Ferrari\\
Departamento de Estatística/IME\\
Universidade de São Paulo\\
Rua do Matão 1010, São Paulo/SP 05508-090, Brazil.\\
{\tt e-mail: silviaferrari.usp@gmail.com} \\
}
\date{}

\begin{document}

\maketitle


\centerline{\rule{14cm}{0.5pt}}
\resumen{
\noindent This paper proposes a general class of regression models for continuous proportions when
the data contain zeros or ones. The proposed class of models assumes that the
response variable has a mixed continuous-discrete distribution with probability mass at zero or one.
The beta distribution is used to describe the continuous component of the model, since its density has
a wide range of different shapes depending on the values of the two parameters that index the distribution.
We use a suitable parameterization of the beta law in terms of its mean and a precision parameter.
The parameters of the mixture distribution are modeled as functions of regression parameters. 
We provide inference, diagnostic, and model selection tools for this class of models. A practical application that employs real data is presented. 
}
{Continuous proportions; Zero-or-one inflated beta distribution; Fractional data; Maximum likelihood estimation; Diagnostics; Residuals.}

\centerline{\rule{14cm}{0.5pt}}

\section{Introduction}

Statistical modeling of continuous proportions has received close attention in the last
few years. Some examples of proportions measured on a continuous scale include 
the fraction of income contributed to a retirement fund, the proportion of weekly hours spent on 
work-related activities, the fraction of household income spent on food, the percentage of 
ammonia escaping unconverted from an oxidation plant, etc. Usual regression models, such as 
normal linear or nonlinear regression models, are not suitable for such situations. Different strategies 
have been proposed for modeling continuous proportions that are perceived to be related to other variables. 
Regression models that assume a beta distribution for the response variable are of particular interest.
 
It is well known that the beta distribution is very flexible for modeling limited range data, since its 
density has different shapes depending on the values of  the two parameters that index the distribution: 
left-skewed, right-skewed, ``$U$," ``$J$," inverted ``$J$," and uniform (see Johnson, Kotz \& Balakrishnan,~1995, 
\S 25.1). Beta regression models have been studied by Kieschnick \& McCullough~(2003), 
Ferrari \& Cribari-Neto~(2004), Espinheira, Ferrari \& Cribari-Neto~(2008a, 2008b), Paolino~(2001), Smithson \& 
Verkuilen~(2006), Korhonen {\it et al}.~(2007), Simas, Barreto-Souza \& Rocha~(2010), and Ferrari \& Pinheiro~(2010), among others. 

Oftentimes, proportions data include a non-negligeable number of zeros and/or ones. 
When this is the case, the beta distribution does not provide a satisfactory description of the data, since it does not
allow a positive probability for any particular point in the 
interval $[0,1]$. A mixed continuous-discrete distribution
might be a better choice. This approach has been considered by Ospina \& Ferrari~(2010), 
who used the beta law to define the continuous component of the distribution. 
The discrete component is defined by a Bernoulli or a degenerate distribution at zero or at one. 
The proposed distributions are usually referred to as zero-and-one inflated beta distributions 
(mixture of a beta and a Bernoulli distribution) and zero-inflated beta distributions or 
one-inflated beta distributions (mixture of a beta and 
a degenerate distribution at zero or at one, depending on the case).
 
This paper is concerned with a general class of regression models for modeling 
continuous proportions when zeros or ones appear in the data. Such class of models is 
tailored for the situation where only one of the extremes (zero or one) is present in the dataset.
We shall consider a mixture of  a beta distribution and a degenerate distribution at a fixed 
known point $c$, where $c \in \{0,1\}$.
The beta distribution is conveniently parameterized in terms of its mean and a precision parameter.
We shall allow the mean and the precision parameter of the beta distribution and the probability of
a point mass at $c$ to be related to linear or non-linear predictors through smooth link functions. 
Inference, diagnostic, and selection tools for the proposed class of models will be presented. 
Closely related to our work are the papers by Hoff~(2007) and Cook, Kieschnick \& McCullough~(2008). 
Our model, however, is more general and convenient than those proposed by these authors. It relaxes 
linearity assumptions, allows all the parameters of the underlying distribution to be modeled 
as functions of unknown parameters and covariates, and uses a suitable parameterization of the beta law.
Unlike their papers, our article offers a comprehensive framework for the statistical analysis of continuous data  observed 
on the standard unit interval with a point mass at one of its extremes.

The paper unfolds as follows. Section 2 presents a general class of zero-or-one inflated beta regression models.
Section 3 discusses maximum likelihood estimation. Section 4 is devoted to diagnostic measures.  Section 5 contains an application using real data and concluding remarks are given in Section 6. Some technical details are 
collected in two appendices.

\section{Zero-or-one inflated beta regression models}

\noindent The beta distribution with parameters $\mu$ and $\phi$ ($0 <\mu <1 $ and $\phi>0$), 
denoted by ${\cal B}(\mu,\phi)$, has the density function
\begin{equation}  
\label{beta}  
f(y;\mu,\phi)=\frac{\Gamma (\phi)}{\Gamma(\mu\phi)\Gamma((1-\mu)\phi)} \  
y^{\mu\phi-1}(1-y)^{(1-\mu)\phi-1},\quad y\in (0,1),  
\end{equation} where $\Gamma(\cdot)$ is the gamma function. 
The parameterization employed in $(\ref{beta})$ is not the usual one, but it is suitable for modeling purposes.

If $y\sim{\cal B}(\mu,\phi)$, then ${\rm E}(y)=\mu$ and 
${\rm Var}(y)=\mu(1 \, - \,\mu)/(\phi \, + \, 1).$ Hence, $\mu$ is the distribution mean and $\phi$ plays the role 
of a precision parameter, in the sense that, for a fixed $\mu$, the larger the value of $\phi,$ the smaller the 
variance of $y.$ Since the beta distribution is very flexible, allowing a wide range of different forms, it is 
an attractive choice for modeling continuous proportions. A possible shortcoming is that
it is not appropriate for modeling datasets that contain observations at the extremes
(either zero or one). Our focus is on the case where only one of the extremes appears in the data---a situation often found in empirical research.
It is then natural to model the data using a mixture of two distributions: a beta
distribution and a degenerate distribution in a known value $c$, where $c$ equals zero 
or one, depending on the case. Under this approach, we assume that 
the probability density function of the 
response variable $y$ with respect to the measure generated by the mixture\footnote{The probability 
measure $P,$ corresponding to this distribution, defined over the measure space 
$((0,1)\cup\{c\},\mathfrak{B}),$ where $\mathfrak{B}$ is the class of all Borelian subsets of  
$(0,1)\cup\{c\},$ is such that $P << \lambda+\delta_{c}$, with $\lambda$ representing the Lebesgue 
measure, and $\delta_{c}$ is such that $\delta_{c}(A)=1$ if $c\in A$ and $\delta_{c}(A)=0$ 
if  $c\notin A$ and $A\in\mathfrak{B}.$} 
is given by 
\begin{equation}  
\label{betamist}  
{\rm bi}_{c}(y;\alpha, \mu,\phi)=
\begin{cases}  
\alpha, & \text{if $y=c $}, \\  
(1-\alpha)f(y; \mu,\phi), & \text{if $y\in(0,1) $},  
\end{cases}  
\end{equation}
where  $f(y; \mu,\phi)$ is the  beta density \eqref{beta}.
Note that $\alpha$ is the probability mass at $c$ and represents the probability of
observing zero $(c=0)$ or one $(c=1)$. 
If $c = 0$, the density \eqref{betamist} is called a zero-inflated beta distribution, and if $c = 1$, the density is called a one-inflated beta distribution.

The mean of $y$ and its variance can be written as
\begin{equation*}  
\label{mediavar}
\begin{aligned}
{\rm E}(y)&=\alpha c+(1-\alpha)\mu, \\
{\rm Var}(y)&=(1-\alpha)\frac{\mu(1-\mu)} {\phi+1}+\alpha(1-\alpha)(c-\mu)^2. \\  
\end{aligned}
\end{equation*}
Note that ${\rm E}(y)$ is the weighted average of the mean of the degenerate distribution
at $c$ and the mean of the beta distribution ${\cal B}(\mu,\phi)$
with weights $\alpha$ and $1-\alpha$. Also,
${\rm E}(y \, |\, y \in (0,1))=\mu$ and ${\rm Var}(y \, |\, y \in (0,1))=\mu(1\,-\,\mu)/(1\,+\,\phi)$. 
Other properties of this distribution can be found in Ospina \& Ferrari~(2010).

A general class of zero-or-one inflated beta regression models  can be defined as follows.
Let $y_1, \ldots, y_n$ be independent random variables such that 
each $y_t$, for $t=1,\ldots,n$, has probability density function  
\eqref{betamist} with parameters $\alpha=\alpha_t$,
$\mu=\mu_t,$ and $\phi=\phi_t$. 
We assume that $\alpha_t$, $\mu_t,$ and $\phi_t$ are defined as
\begin{equation}
\label{relacao}
\begin{aligned}
h_1(\alpha_t) &=\eta_{1t}=f_1(v_t,\rho),\\
h_2(\mu_t)    &=\eta_{2t}=f_2(x_t,\beta),\\
h_3(\phi_t)   &=\eta_{3t}=f_3(z_t,\gamma),\\
\end{aligned}
\end{equation} 
where $\rho=(\rho_1,\ldots,\rho_p)^\top$, $\beta=(\beta_1,\ldots,\beta_k)^\top$ and $\gamma=(\gamma_1,\ldots,\gamma_m)^\top$ 
are vectors of unknown regression parameters; ($p\,+\,k\,+\,m<n$),
$\eta_1=(\eta_{11}, \ldots, \eta_{1n})^\top$, $\eta_2=(\eta_{21}, \ldots, \eta_{2n})^\top,$ and
$\eta_3=(\eta_{31}, \ldots, \eta_{3n})^\top$ are 
predictor vectors; and $f_1(\cdot,\cdot),$ $f_2(\cdot,\cdot),$ and $f_3(\cdot,\cdot)$ are linear or nonlinear 
twice continuously differentiable functions in the second argument, such that the derivative matrices 
${\cal V}=\partial\eta_1/\partial \rho,$ ${\cal X}=\partial\eta_2/\partial \beta,$ and ${\cal Z}=\partial\eta_3/\partial \gamma$
have ranks $p,$ $k$ and $m$, respectively, for all $\rho,$ $\beta,$ and $\gamma.$

Here $v_t=(v_{t1},\ldots,v_{tp'})$, $x_t=(x_{t1},\ldots,x_{tk'}),$ and
$z_t=(z_{t1},\ldots,z_{tm'})$ are observations on $p'+k'+m'$  known covariates. 
We also assume that the link functions $h_1\!\!:(0,1)\rightarrow \mathbb{R}$, $h_2\!\!:(0,1)\rightarrow \mathbb{R}$,
and $h_3\!\!:(0,+\infty)\rightarrow \mathbb{R}$ are strictly monotonic and twice differentiable.
Various different link functions may be used. For $\mu$ and $\alpha$ one may choose 
$h_2(\mu)=\log\{\mu/(1-\mu)\}$ (logit link); $h_2(\mu)=\Phi^{-1}(\mu)$ (probit link), 
where $\Phi(\cdot)$ denotes the standard normal distribution function; $h_3(\mu)=\log\{-\log(1-\mu)\}$ (complementary log-log link); and $h_2(\mu)=-\log\{-\log(\mu)\}$ (log-log link), among others. 
Possible specifications for $\phi$ are $h_3(\phi)=\log \phi$ (log link) and
$h_3(\phi)=\sqrt{\phi}$ (square-root link).

Model (\ref{betamist})-(\ref{relacao})  has a number of interesting features. 
The variance of $y_t$ is a function of $(\alpha_t,\mu_t,\phi_t)$ and, as a consequence, 
of the covariates values. Hence, non-constant response variances are naturally accommodated by 
the model. Also, the role that the covariates and the parameters play in the model is clear. For example,
suppose  $c=0$. In this case, 
$v_t$ and $\rho$ affect ${\rm Pr}(y_t=0)$,  $x_t$ and $\beta$ control ${\rm E}(y_t \, |\, y_t \in (0,1)),$ 
and $z_t$ and $\gamma$ influence the precision of the conditional distribution of $y_t,$ given that
$y_t \in (0,1)$. This feature can be very useful for modeling purposes.
For instance, if the response is the individual mobile communications 
expenditure proportion (MCEP), model (\ref{betamist})-(\ref{relacao})  allows the researcher to take into account that 
some individuals do not spend at all on MCEP and to separately assess the effects of the heterogeneity 
among consumers and non-consumers of mobile communications (in this connection, see Yoo, 2004). 

\paragraph{Special cases.}
Model (\ref{betamist})-(\ref{relacao}) embodies two general classes of models: the {\it zero-inflated beta 
regression model} ($c=0$) and the {\it one-inflated beta regression model} ($c=1$), the first of which is 
suitable when the data include zeros and the second, when ones appear in the dataset.
Each of them leads to a corresponding linear model when the predictors are
linear functions of the parameters. In this case, we have $p=p'$, $k=k'$, $m=m'$, $h_1(\alpha_t)=v_t^\top\rho,$ $h_2(\mu_t)=x_t^\top\beta$ and $h_3(\phi_t)= z_t^\top\gamma.$ Here, ${\cal V} = (v_1^\top,\ldots,v_n^\top)^\top$,
 ${\cal X} =  (x_1^\top,\ldots,x_n^\top)^\top$, and ${\cal Z}  = (z_1^\top,\ldots,z_n^\top)^\top$.
Also, the nonlinear beta regression model (Simas, Barreto-Souza \& Rocha, 2010, and Ferrari \& Pinheiro, 2010) is a limiting case of our model obtained by setting $\alpha_t=\alpha\rightarrow0$.
If, in addition, the predictor for $\mu_t$ is linear and $\phi_t$ is constant through the observations, we
arrive at the beta regression model defined by Ferrari \& Cribari-Neto~(2004).

\section{Likelihood inference}

The likelihood function for $\theta=(\rho^\top,\beta^\top,\gamma^\top)^\top$ 
based on a sample of $n$ independent observations is  
\begin{equation}
\label{verfatorada}
L(\theta)=\prod_{t=1}^n{\rm bi}_c(y_t;\alpha_t,\mu_t,\phi) =L_1(\rho)L_2(\beta,\gamma),
\end{equation}
where
\begin{equation*}
\begin{aligned}
L_1(\rho)&=\prod_{t=1}^n\alpha_t^{{\rm 1\!l}_{\{c\}}(y_t)}(1-\alpha_t)^{1-{\rm 1\!l}_{\{c\}}(y_t)}, \\
L_2(\beta,\gamma)&=\prod_{t:y_t\in(0,1)}f(y_t;\mu_t,\phi_t),  \\
\end{aligned}
\end{equation*}
with ${\rm 1\!l}_{A}(y_t)$ being an indicator function that equals 1 if $y_t\in A,$ and  0, if $y_t\notin A.$ Here, $\alpha_t=h_1^{-1}(\eta_{1t}),$ $\mu_t=h_2^{-1}(\eta_{2t}),$ and $\phi_t=h_3^{-1}(\eta_{3t})$, as defined in \eqref{relacao}, are  functions of $\rho,$ $\beta,$ and $\gamma$, respectively. Notice that the likelihood function $L(\theta)$ factorizes in two terms, 
the first of which depends only on $\rho$ (discrete component),
and the second, only on $(\beta,\gamma)$ (continuous component). Hence, the parameters are separable (Pace \& Salvan, 1997, p.\ 128) and the maximum likelihood inference for $(\beta,\gamma)$ can be performed separately from that for $\rho,$ as if the value of $\rho$ were known, and vice-versa. 

The log-likelihood function is given by
\begin{equation}
\label{loglik}
\ell(\theta)=\ell_1(\rho)+\ell_2(\beta,\gamma)=\SumD\ell_t(\alpha_t)+\!\!\!\! \sum_{t:y_t\in(0,1)}\ell_t(\mu_t, \phi_t),
\end{equation}
where
\begin{equation}
\label{logrho}
        \ell_t(\alpha_t)= {{\rm 1\!l}_{\{c\}}(y_t)}\log\alpha_t +(1-{{\rm 1\!l}_{\{c\}}(y_t)})\log(1-\alpha_t),
\end{equation} 
\begin{equation}
\label{logbeta}
        \ell_t(\mu_t, \phi_t)=\log\Gamma(\phi_t)-\log\Gamma(\mu_t\phi_t)-\log\Gamma((1-\mu_t)\phi_t)+(\mu_t\phi_t-1)y_{t}^{\ast}
        +(\phi_t-2)y_t^\dagger,
\end{equation} 
$y_{t}^{\ast}=\log\{y_t/(1-y_t)\}$ and $y_{t}^{\dagger}=\log(1-y_t)$  if $y_t\in(0,1),$ and $y_{t}^{\ast}=0$
and $y_{t}^{\dagger}=0$ otherwise. 
We have the following conditional moments of $y_t^*$ and $y_t^\dagger$: 
\begin{equation}
\label{mom_y}
\begin{aligned}
\mu_{t}^{\ast }&={\rm E}(y_{t}^{\ast}\,|\,y_t\in (0,1))=\psi(\mu _{t}\phi_t)-\psi(( 1-\mu _{t})\phi_t), \\
\mu_{t}^\dagger &={\rm E}(y_{t}^{\dagger}\,|\,y_t\in (0,1))=\psi(( 1-\mu _{t})\phi_t)-\psi(\phi_t), \\
v_{t}^{\ast } & = {\rm Var}(y_{t}^{\ast}\,|\,y_t\in (0,1)) = \psi'(\mu _{t}\phi_t)-\psi'(( 1-\mu _{t})\phi_t), \\ 
v_{t}^\dagger &  ={\rm Var}(y_{t}^{\dagger}\,|\,y_t\in (0,1)) = \psi'(( 1-\mu _{t})\phi_t)-\psi'(\phi_t), \\
c^{\ast\dagger} & = {\rm Cov}(y_{t}^{\ast}, y_{t}^{\dagger}\,|\,y_t\in (0,1))=-\psi'(( 1-\mu _{t})\phi_t),
\end{aligned}
\end{equation}
with $\psi(\cdot)$ denoting the digamma function.\footnote{\noindent\small The digamma function is defined 
as $\psi(x)={\rm d} \log\Gamma(x)/{\rm d}x, \ x>0.$ 
} 
Notice that $\ell_1(\rho)$ represents the log-likelihood function of a regression model for binary responses, in which the success probability for the $t$th observation is $\alpha_t=h_1^{-1}(\eta_{1t})$ (McCullagh \& Nelder~1989, \S 4.4.1). On the other hand, $\ell_2(\beta,\gamma)$ is the log-likelihood function for $(\beta,\gamma)$ in a nonlinear
beta regression model based on the observations that fall in the interval $(0,1)$.
Hence, the maximum likelihood (ML) estimation for this model can be accomplished by separately fitting a binomial regression model to the indicator variables ${{\rm 1\!l}_{\{c\}}(y_t)},$ for $t=1,\ldots,n,$ and a nonlinear beta
regression model to the observations $y_t\in(0,1)$, for $t=1,\ldots,n$. 

The score function, obtained by the differentiation of the log-likelihood function with respect to the unknown parameters (see \eqref{Us} -- \eqref{UR}; Appendix A), is given by $$U(\theta)= (U_{\rho}(\rho)^\top, U_{\beta}(\beta, \gamma)^\top, U_{\gamma}(\beta, \gamma)^\top)^\top,$$ where
\begin{equation}
\label{scoref}
\begin{aligned}
U_{\rho}(\rho)  & =  {\cal V}^\top {\cal A} { D} {\cal A}^*(y^c - \alpha),  \\
U_{\beta}(\beta, \gamma) &= {\cal X}^\top (I_n-Y^c) T \Phi(y^* -\mu^*), \\
U_{\gamma}(\beta, \gamma) & = {\cal Z}^\top (I_n-Y^c) H [{\cal M}(y^*-\mu^*)+(y^\dagger-\mu^\dagger)]. \\
\end{aligned}
\end{equation} 
Here,  
$y^*=(y_1^*,\ldots,y_n^*)^\top,$ \, 
$y^\dagger=(y_1^\dagger,\ldots,y_n^\dagger)^\top,$ \,
$y^c=({{\rm 1\!l}_{\{c\}}(y_1)},\ldots,{{\rm 1\!l}_{\{c\}}(y_n)})^\top,$ \,
$\mu^*=(\mu_1^*,\ldots,\mu_n^*)^\top,$ \,
$\mu^\dagger=(\mu_1^\dagger,\ldots,\mu_n^\dagger)^\top,$ and
$\alpha=(\alpha_1,\ldots,\alpha_n)^\top$ are $n$-vectors and 
${\cal M} ={\rm diag}(\mu_1,\ldots, \mu_n),$ \,
${\cal A} ={\rm diag}(1/\alpha_1, \ldots, 1/\alpha_n),$ \,
${\cal A}^* ={\rm diag}(1/(1-\alpha_1), \ldots, 1/(1-\alpha_n)),$ \,
${ D}={\rm diag}({1}/{h'_1(\alpha_1)}, \break\hfil \ldots,  {1}/{h'_1(\alpha_n)}),$ \,
$\Phi={\rm diag}(\phi_1,\ldots,\phi_n),$ \,
$T = {\rm diag}({1}/{h'_2(\mu_1)}, \ldots, {1}/{h'_2(\mu_n)}),$ \,
$H={\rm diag}({1}/{h'_3(\phi_1)},  \break\hfil \ldots,   {1}/{h'_3(\phi_n)}),$  and
$Y^c={\rm diag}({{\rm 1\!l}_{\{c\}}(y_1)},\ldots,{{\rm 1\!l}_{\{c\}}(y_n)})$  
are $n \times n$ diagonal matrices. Also, $I_n$ represents the $n\times n$ identity matrix.
The maximum likelihood estimators (MLEs) of $\rho$ and $(\beta^\top,\gamma^\top)^\top$ are obtained as the solutions of the nonlinear systems $U_\rho(\rho)=0$ and $(U_\beta(\beta,\gamma)^\top, U_\phi(\beta,\gamma))^\top=0.$ There are not closed form expressions for these estimators, and their computations should be performed numerically using a nonlinear optimization algorithm, e.g., some form of Newton's method  (Newton-Raphson, Fisher's scoring, BHHH, etc.) or a quasi-Newton algorithm such as BFGS. An iterative algorithm for maximum likelihood estimation is presented in Appendix \ref{AppB}. For more details on nonlinear optimization, see Press {\it et al.}~(1992). 
 
From the observed information matrix given in  \eqref{J}, it can be shown that the Fisher information matrix is 
\begin{equation}
\label{fisher}
 K(\theta) = 
\begin{pmatrix}
K_{\rho\rho} & 0 & 0 \\
0            &  K_{\beta\beta} &  K_{\beta\gamma} \\
0            &  K_{\gamma\beta} &  K_{\gamma\gamma} \\
\end{pmatrix},
\end{equation}  
where
$ K_{\rho\rho}     =  {\cal V}^\top \mathbf{{W_1}} {\cal V}, $ 
$ K_{\beta\beta}   = {\cal X}^\top \mathbf{{W_2}}  {\cal X},$
$ K_{\gamma\beta} = {K^\top_{\beta\gamma}} = {\cal X}^\top \mathbf{{W_3}} {\cal Z},$ $K_{\gamma\gamma} =   {\cal Z}^\top \mathbf{{W_4}} {\cal Z}$ 
with
$ \mathbf{{W_1}}    = ({\cal A}^{*} +{\cal A}) {\cal D}^2, $ $  \mathbf{{W_2}}  =  \Phi T \{ V^* {\cal A}^{* -1}\} T \Phi,$
$ \mathbf{{W_3}} = T \{\Phi({\cal M} V^* + C ){\cal A}^{* -1}\}   H,$
and  $\mathbf{{W_4}} =    H\{  ({\cal M}^2  V^* +2{\cal M} C + V^\dagger) {\cal A}^{* -1}\} H .$ Notice that  $K_{\rho\beta}={K_{\beta\rho}}^\top= 0$ and $K_{\rho\gamma}={K_{\gamma\rho}}^\top= 0$, 
thus indicating that the parameters $\gamma$ and $(\beta^\top,\gamma^\top)^\top$ are globally orthogonal 
(Cox \& Reid, 1987) and their MLEs,  $\widehat\rho$ and
$(\widehat\beta^\top,\widehat\gamma^\top)^\top,$ are asymptotically independent. 

The inverse of Fisher's information matrix is useful for computing asymptotic standard errors of MLEs. 
From \eqref{fisher} and a standard formula for the inverse of partitioned matrices (Rao, 1973, p.\ 33), we have
 \begin{equation}
\label{invfisher}
 K(\theta)^{-1} = 
\begin{pmatrix}
K^{\rho\rho} & 0 & 0 \\
0            &  K^{\beta\beta} &  K^{\beta\gamma} \\
0            &  K^{\gamma\beta} &  K^{\gamma\gamma} \\
\end{pmatrix},
\end{equation} 
where  
$ K^{\rho\rho}     =  ({\cal V}^\top \mathbf{{W_1}} {\cal V})^{-1},$ 
$ K^{\beta\beta}   = 
\{{\cal X}^\top( \mathbf{{W_2}}   -  \mathbf{{W_3}} {\cal Z}({\cal Z}^\top \mathbf{{W_4}} {\cal Z})^{-1} 
{\cal Z}^\top \mathbf{{W_3}} ){\cal X}\}^{-1},
$ 
$ K^{\gamma\beta} = {(K^{\beta\gamma})^\top } = 
-({\cal Z}^\top \mathbf{{W_4}} {\cal Z})^{-1} {\cal Z}^\top \mathbf{{W_3}} {\cal X}({\cal X}^\top( \mathbf{{W_2}}   -  \mathbf{{W_3}} {\cal Z}({\cal Z}^\top \mathbf{{W_4}} {\cal Z})^{-1} 
{\cal Z}^\top \mathbf{{W_3}} ){\cal X})^{-1},
$ 
and $K^{\gamma\gamma} \break =   ({\cal Z}^\top \mathbf{{W_4}} {\cal Z})^{-1} + 
({\cal Z}^\top \mathbf{{W_4}} {\cal Z})^{-1}
{\cal Z}^\top \mathbf{{W_3}} {\cal X}
({\cal X}^\top( \mathbf{{W_2}}   -  \mathbf{{W_3}} {\cal Z}({\cal Z}^\top \mathbf{{W_4}} {\cal Z})^{-1} 
{\cal Z}^\top \mathbf{{W_3}} ){\cal X})^{-1} \break\hfill
{\cal X}^\top \mathbf{{W_3}} {\cal Z}
\times
({\cal Z}^\top \mathbf{{W_4}} {\cal Z})^{-1}.
$ 

\paragraph{Fitting the model using the GAMLSS implementation.}
The zero-or-one inflated beta distribution has been incorporated in the GAMLSS framework (Rigby \& Stasinopoulos, 2005);
see Ospina (2006). GAMLSS allows the flexible modeling of each of the three parameters that index the distribution using parametric terms involving linear or nonlinear predictors, smooth nonparametric terms (e.g., cubic splines or loess), and random effects. Maximum (penalized) likelihood estimation is approached through a Newton-Raphson or Fisher scoring algorithm with the backfitting algorithm for the additive components. Our approach consists of an application of the {\tt gamlss} functions, which are fully documented in the {\tt gamlss} package (Stasinopoulos \& Rigby, 2007; see also {\tt http://www.gamlss.org}). The structure of the {\tt gamlss} functions is familiar to readers who are used to the {\tt R} (or {\tt S-Plus}) syntax (the {\tt glm} function, in particular). The set of {\tt gamlss} packages can be freely downloaded from  the {\tt R} library at {\tt http://www.r-project.org/}.

\paragraph{Large sample inference.} 
If the model specified by \eqref{betamist} and \eqref{relacao} is valid and the usual regularity conditions 
for maximum likelihood estimation are satisfied (Cox \& Hinkley, 1974, p.\ 107),
the MLEs of $\theta$ and $K(\theta)$, $\widehat\theta=(\widehat\rho^\top \widehat\beta^\top,\widehat\gamma)^\top$
and $K(\widehat\theta),$ respectively, are consistent.
Assuming that $I(\theta)=\lim_{n \rightarrow \infty} \{n^{-1}K(\theta)\}$ exists and is nonsingular,
we have  $\sqrt{n}(\widehat\theta - \theta) \ {\buildrel {\cal D} \over \rightarrow
\ {\cal N}_{p+k+m}}(0,I(\theta)^{-1}),$ where ${\buildrel {\cal D} \over \rightarrow }$ denotes convergence in  distribution. Note that, if $\theta_l$ denotes the $l$th component of $\theta$, it follows that 
$$
(\widehat\theta_l - \theta_l)\left\{K(\widehat\theta)^{ll}\right\}^{-1/2}\ \
{\buildrel {\cal D} \over \rightarrow \ \ {\cal N}}(0,1),
$$
where $K(\widehat\theta)^{ll}$ is the $l$th diagonal element of
$K(\widehat\theta)^{-1}.$ A rigorous proof of the aforementioned asymptotic result can be obtained by extending arguments similar to those given in Fahrmeir \& Kaufmann (1985).

Let $K(\widehat\rho)^{ss}$, $K(\widehat\beta)^{rr},$ and $K(\widehat\gamma)^{RR}$ be the estimated
$s$th, $r$th and $R$th diagonal elements of $K^{\rho\rho}$, $K^{\beta\beta},$ and $K^{\gamma\gamma}$,
respectively. The asymptotic variances of $\,\widehat{\!\rho}_s,$ $\,\widehat{\!\beta}_r,$ and $\,\widehat{\!\gamma}_R$  are estimated by $K(\widehat\rho)^{ss},$ $K(\widehat\beta)^{rr},$ 
and $K(\widehat\gamma)^{RR},$ respectively. If $0<\varsigma < 1/2$, and $z_\varsigma $
represents  the $\varsigma$th quantile of the ${\cal N}(0,1)$ distribution, we have 
$\,\widehat{\!\rho}_s \pm z_{1-{\varsigma/2}}(K(\widehat\rho)^{ss})^{1/2},$
$\,\widehat{\!\beta}_r \pm z_{1-{\varsigma/2}}(K(\widehat\beta)^{rr})^{1/2},$ 
and $\,\widehat{\!\gamma}_R \pm z_{1-{\varsigma/2}}(K(\widehat\gamma)^{RR})^{1/2}$
as the limits of asymptotic confidence intervals (ACI) for $\rho_s,$ $\beta_r,$ and $\gamma_R$, 
respectively, all with asymptotic coverage $100(1-\varsigma)$\%. 
Additionally,  an approximate $100(1-\varsigma)$\% confidence interval for ${\mu^\bullet}={\rm E}(y)$,
the mean response for the given covariates $v$ and $x,$ can be computed as 
$$
\Bigr(
        \widehat{{\mu^\bullet}}-z_{_{1-{\varsigma/2}}} \ {\rm s.e.}(\widehat{\mu^\bullet}),  \ \ 
        \widehat{{\mu^\bullet}}+z_{_{1-{\varsigma/2}}}\ {\rm s.e.}(\widehat{{\mu^\bullet}})
\Bigl),
$$ where $c=0$ or $c=1,$ depending on the case;
$\widehat{{\mu^\bullet}} =c\widehat\alpha+(1-\widehat\alpha)\widehat\mu,$ 
with $\widehat\alpha =h_1^{-1}(\widehat\eta_{1});$ $\widehat\mu =h_2^{-1}(\widehat\eta_{2});$
$\widehat\eta_{1} = {\textit{f}}_{ 1}( v;\widehat\rho);$
$\widehat\eta_{2} = {\textit{f}}_{ 2}( x;\widehat\beta);$ and 
$$
{\rm s.e.}(\widehat{\mu^\bullet})=\sqrt
{
    \big\{{(c-\widehat\mu)/{h_1'(\widehat\alpha)}}\big\}^2v^\top 
    \widehat K^{\rho\rho}v+\big\{{(1-\widehat\alpha)/
    {h_2'(\widehat\mu)}}\big\}^2x^\top \widehat K^{\beta\beta}x
}.
$$ 
Here, $\widehat K^{\rho\rho}$ and $\widehat K^{\beta\beta}$ respectively equal $K^{\rho\rho}$ and $K^{\beta\beta},$ evaluated at $\widehat \theta$.

The likelihood ratio,  Rao's score, and Wald's (W) statistics 
to test hypotheses on the parameters can be calculated from the log-likelihood function, the score vector,
the Fisher information matrix, and its inverse given above. Their null distributions are usually unknown 
and the tests rely on asymptotic approximations. In large samples, a chi-squared distribution can be used as an  approximation to the true null distributions. For testing the significance of the $i$th regression parameter that models $\mu,$  one can use the signed square root of Wald's statistic, $\widehat\beta_i / {\rm s.e.}(\widehat\beta_i),$ where ${\rm s.e.}(\widehat\beta_i)$ is the asymptotic standard error of the MLE of 
$\beta_i$ obtained from the inverse of Fisher's information matrix evaluated at the maximum likelihood
estimates. The limiting null distribution of the test statistic is standard normal. Significance tests
on the $\gamma$'s and $\rho$'s can be performed in a similar fashion.

\paragraph{Finite-sample performance of the MLEs.}
We now present the results of two Monte Carlo simulation experiments in order to investigate the finite-sample performance of the MLEs. The first experiment evaluates the impact of the magnitude of the probability of zero response on the MLEs. Here, the sample size is $n=150$ and we focus on the zero-inflated beta regression model with ${\rm logit}(\mu)    =\beta_0 + \beta_1 x_1 + \beta_2 x_2 +\beta_3 x_3,$ $\log(\phi)   =\gamma_0 + \gamma_1 z_1 + \gamma_2 z_2 +\gamma_3 z_3,$  and $\alpha$ being constant for all observations.  
The true parameter values were taken as 
$\beta_0 =-1$, 
$\beta_1 = 1$, 
$\beta_2 = -0.5$, 
$\beta_3 = 0.5$, 
$\gamma_0 = 2$, 
$\gamma_1 = 1$, 
$\gamma_2 = 0.5,$
and 
$\gamma_3 = 0.5.$  
Here, we consider four different values for the probability of observing zero: $\alpha=0.18,$ $\alpha=0.32,$ $\alpha=0.68,$ and $\alpha=0.82.$

The second experiment considers the zero-inflated beta regression model with
${\rm logit}(\alpha) =\rho_0 + \rho_1 v_1 + \rho_2 v_2 +\rho_3 v_3,$ 
$ 
{\rm logit}(\mu)  =\beta_0 + \beta_1 x_1 + \beta_2 x_2 +\beta_3 x_3,$ and $\log(\phi)   =\gamma_0 + \gamma_1 z_1 + \gamma_2 z_2 +\gamma_3 z_3.$ 
Here, we evaluate the performance of the MLEs when the sample size increases. The true values of the $\beta$'s and the $\gamma$'s are the same as in the first experiment. The true values of the $\rho$'s are
$\rho_0 =-1$, 
$\rho_1 = 1$, 
$\rho_2 = -0.5,$ and
$\rho_3 = 0.5.$ 
In this situation the sample sizes considered are 
$n=50, 150,$ and $300.$ 

The  explanatory variables $v_1, x_1,$ and $z_1$ were generated as independent draws from a standard normal distribution. 
The covariates $v_2, x_2,$ and $z_2$ were generated from the Poisson distribution with unit mean, and $v_3, x_3,$ and 
$z_3$ were generated from the ${\rm Binomial}(0.2, 5)$ distribution. The total number of Monte Carlo replications was set at 5000 for each sample size. 
All simulations were carried out in {\tt R} (Ihaka \& Gentleman,~1996). Computations for fitting inflated beta regression models were performed using the {\tt gamlss} package. The MLEs were obtained by maximizing the log-likelihood function using the {\tt RS} algorithm (Rigby  \& Stasinopoulos, 2005). In order to analyze the results, we computed the bias and the root mean squared error ($\sqrt{{\rm MSE}}$) of the estimates.

Table \ref{tab:n1} summarizes the numerical results of the first experiment. 
We note that, 
for a fixed sample size, the bias and the $\sqrt{{\rm MSE}}$ of the estimators of the continuous component ($\beta$'s and $\gamma$'s) increase with the expected proportion of zeros ($\alpha$). 
This is to be expected, since the expected number of observations in $(0,1)$ decreases as $\alpha$ increases.
Also, it is noteworthy that the bias and the $\sqrt{{\rm MSE}}$ of MLEs corresponding to the dispersion covariates tend to be much more pronounced when compared with the MLEs of the parameters that model the mean response.

\begin{table}[!ht]
\begin{center}
\caption{Simulation results for the first experiment.}
\label{tab:n1}
\medskip
{\footnotesize
\renewcommand{\tabcolsep}{0.6pc} 
\renewcommand{\arraystretch}{1.4} 
\begin{tabular}{@{}crrrrrrrr}\hline\hline\noalign{\smallskip}
& \multicolumn{2}{c}{$\alpha =  0.18$}  & \multicolumn{2}{c}{$\alpha =  0.18$} & \multicolumn{2}{c}{$\alpha =  0.18$} & \multicolumn{2}{c}{$\alpha =  0.18$}  
\\ 
\cmidrule(r){2-3} 
\cmidrule(r){4-5}
\cmidrule(r){6-7}  
\cmidrule(r){8-9} 

Estimator & Bias & $\sqrt{{\rm MSE}}$ & Bias & $\sqrt{{\rm MSE}}$ & Bias & $\sqrt{{\rm MSE}}$ & Bias & $\sqrt{{\rm MSE}}$ \\ \hline\hline

$\widehat\alpha$  & $  0.00052 $&$ 0.03149 $ & $  0.00022 $&$ 0.03778 $  & $  0.00086 $&$ 0.03802 $ & $ -0.00207 $&$ 0.03097 $ \\ \hline\hline      

$\widehat\beta_0$ & $  -0.00095$&$  0.05213$ & $  -0.00176$&$  0.07724$ &  $   0.00032$&$  0.11219$ &  $  -0.00226$&$  0.19777$ \\
$\widehat\beta_1$ & $   0.00176$&$  0.04006$&  $   0.00207$&$  0.04101$&  $   0.00354$&$  0.07070$ &  $   0.00452$&$  0.11554$ \\
$\widehat\beta_2$ &  $  -0.00048$&$  0.02711$ &  $  -0.00076$&$  0.04308$&  $  -0.00427$&$  0.08532$&  $  -0.00328$&$  0.11987$ \\
$\widehat\beta_3$ &  $   0.00014$&$  0.03764$&  $   0.00024$&$  0.05145$&  $   0.00073$&$  0.08256$&  $   0.00089$&$  0.12046$ \\  \hline\hline      

$\widehat\gamma_0$&  $   0.01241  $&$ 0.21359 $&   $   0.02668  $&$ 0.25779 $&   $   0.02892  $&$ 0.43969 $&   $   0.06437  $&$ 0.71228 $\\ 
$\widehat\gamma_1$&  $   0.03158  $&$ 0.13205 $&   $   0.03203  $&$ 0.14619 $&   $   0.07302  $&$ 0.30639 $&   $   0.20369  $&$ 0.51323 $ \\
$\widehat\gamma_2$ &   $   0.03088  $&$ 0.12514 $&   $   0.04288  $&$ 0.16232 $&   $   0.09854  $&$ 0.30874 $&   $   0.23140  $&$ 0.68394 $ \\
$\widehat\gamma_3$&   $   0.02745  $&$ 0.15964 $&   $   0.01527  $&$ 0.17183 $&   $   0.07067  $&$ 0.32288 $&   $   0.19891  $&$ 0.69026 $ \\  \hline\hline

    \end{tabular}
}
\end{center}    
\end{table}

Table \ref{tab:s1} presents simulation results for the second experiment.  For the smallest sample size considered ($n=50$), the estimation algorithm failed to converge in 1.3\% of the samples. 
For large sample sizes the algorithm converged for all the samples. 
The estimated biases of the MLEs  of the $\rho$'s (parameters of the discrete component)  are markedly high for small samples. This is not surprising; in standard logistic regression, the MLEs are
considerably biased in small samples. For large samples, the bias of the $\rho$'s is negligible.

In the second experiment, the $\beta$'s and the $\gamma$'s are essentially estimated from the observations in (0,1), which represent around 70\% of the observations in our study.  For all the sample sizes considered,  the mean of the $\widehat\beta$'s  and $\widehat\gamma$'s are close to the corresponding true values. Also, all the root mean square errors decrease when the  sample size increases, as expected.

\begin{table}[!ht]
\begin{center}
\caption{Simulation results for the second experiment.}
\label{tab:s1}
\medskip
{\footnotesize
\renewcommand{\tabcolsep}{0.8pc} 
\renewcommand{\arraystretch}{1.3} 
\begin{tabular}{@{}crrrrrr}\hline\hline\noalign{\smallskip}

& \multicolumn{2}{c}{$n =  50$}  & \multicolumn{2}{c}{$n =  150$} & \multicolumn{2}{c}{$n =  300$} 
\\ 
\cmidrule(r){2-3} 
\cmidrule(r){4-5}
\cmidrule(r){6-7}  

Estimator & Bias & $\sqrt{{\rm MSE}}$ & Bias & $\sqrt{{\rm MSE}}$ & Bias & $\sqrt{{\rm MSE}}$  \\ \hline \hline

$\widehat\rho_0$ &$-0.14554    $&$ 0.81179   $ &$-0.02398     $&$ 0.38136   $&$-0.01263     $&$ 0.26529   $ \\          
                                                       
$\widehat\rho_1$&$-0.18009     $&$ 0.56946   $ &$-0.04623     $&$ 0.25131   $&$-0.02501     $&$ 0.17076   $ \\
                                                 
$\widehat\rho_2$&$-0.13892     $&$ 0.66011   $ &$-0.03437     $&$ 0.24178   $&$-0.01803     $&$ 0.16958   $ \\
                                                 
$\widehat\rho_3$&$ 0.09667     $&$ 0.55895   $ &$ 0.02113     $&$ 0.24240   $&$ 0.01093     $&$ 0.16877   $ \\   \hline \hline

$\widehat\beta_0$&$ 0.00588   $&$0.16190  $ &$ 0.00171   $&$0.07288  $ &$ 0.00055   $&$0.05272  $ \\
                                                 
$\widehat\beta_1$&$ 0.00588   $&$0.10917  $ &$ 0.00146   $&$0.04882  $ &$ 0.00027   $&$0.03313  $ \\
                                                 
$\widehat\beta_2$&$-0.00211   $&$0.09761  $ &$-0.00131   $&$0.05057  $ &$ 0.00018   $&$0.02872  $ \\
                                                 
$\widehat\beta_3$&$ 0.00140   $&$0.12356  $ &$ 0.00104   $&$0.05264  $  &$ 0.00063   $&$0.03049  $  \\  \hline \hline
                                                                                                        
$\widehat\gamma_0$&$ 0.12628  $&$ 0.45696   $  &$ 0.04208  $&$ 0.25820   $ &$ 0.02472  $&$ 0.18738   $ \\
                                               
$\widehat\gamma_1$&$ 0.04197  $&$ 0.29273   $  &$ 0.03485  $&$ 0.15767   $ &$ 0.00959  $&$ 0.10359   $  \\
                                               
$\widehat\gamma_2$&$ 0.09771  $&$ 0.39828   $  &$ 0.02333  $&$ 0.17590   $ &$ 0.01153  $&$ 0.10208   $ \\
                                               
$\widehat\gamma_3$ &$ 0.04934  $&$ 0.36516   $ &$ 0.02266  $&$ 0.18366   $ &$ 0.00473  $&$ 0.11669   $   \\  \hline \hline            
    
    \end{tabular}
}
\end{center}    
\end{table}

\section{Diagnostics}

Likelihood-based inference depends on parametric assumptions, and a severe misspecification of the model or the 
presence of outliers may impair its accuracy. We shall introduce some types of residuals for detecting 
departures from the postulated model and outlying  observations. Additionally, we suggest some measures to 
assess  goodness-of-fit.

\paragraph{Residuals.} 
Residual analysis in the context of the zero-or-one inflated beta regression model (\ref{beta})-(\ref{relacao})
can be split into two parts. First, we focus on the residual analysis for the discrete and the continuous component 
of the model separately. For this purpose, we propose a standardized Pearson residual based on Fisher's scoring 
iterative algorithm for estimating $\rho,$ $\beta,$ and $\gamma.$ Second, we define a randomized quantile residual 
as a global residual for the model, i.e., using information of the discrete and the continuous component
simultaneously. 

To study departures from the error assumptions as well as the presence of outlying observations of the discrete 
component for the zero-or-one inflated beta regression model we define a version of the standardized Pearson 
residual as
\begin{equation}
\label{rpd}
r_{pt}^{\rm D} = \frac{{{\rm 1\!l}_{\{c\}}(y_t)} - \widehat\alpha_t}{\sqrt{ {\widehat\alpha_t(1-\widehat\alpha_t})(1-\widehat {\mathbf{h}}_{tt}) }  }, \qquad t=1,\ldots,n,
\end{equation}
where $\mathbf{h}_{tt}$ is the $t$th diagonal element of the orthogonal projection matrix $\mathbf{{H}}$ 
defined in \eqref{hm} and ``\,$ \ \widehat{} \ $\," indicates evaluation at the MLEs. 
Note that $r_{pt}^{\rm D}$ is a version of the ordinary residual obtained in (\ref{rescd}) that takes the leverage of the $t$th observation into account. Plots of the residuals against the  covariates or the fitted values should exhibit no systematic trend. 

The conditional distribution of $y_t,$ given that $y_t \in (0,1),$ is a beta distribution ${\cal B}(\mu_t,\phi_t).$ 
Ferrari and Cribari--Neto~(2004) provided two different residuals (standardized and deviance) for the class of beta regression models. Espinheira {\it et al}.~(2008b) proposed two new beta regression residuals and their numerical results favor one of the new residuals: the one that accounts for the leverages of the different observations. Here,
we follow Espinheira {\it et al.}~(2008b) to define a weighted version of the ordinary residual from Fisher's scoring iterative algorithm for estimating $\beta$ when $\gamma$ is fixed (see Appendix \ref{AppB}). Our proposed residual is defined as
\begin{equation}
\label{rpc}
r_{pt}^{\rm C} =  \frac{y_t^*-\widehat\mu_t^*}{\sqrt{
\widehat{v}^*_{t} (1-\widehat\alpha_t)(1-\widehat{\mathbf{{P}}}_{tt}) }}, \qquad t: y_t \in (0,1),
\end{equation} where $\widehat{\mathbf{{P}}}_{tt}$ is the $t$th diagonal element of the projection matrix
${\mathbf{{\widehat P}}}$ defined in \eqref{hatP}. Note that $r_{pt}^{\rm C}$ is a version of the standardized Pearson residual obtained in \eqref{r} that takes the leverage of the $t$th observation into account.

To assess the overall adequacy of the zero-or-one inflated beta regression model to the data at hand, we propose the randomized quantile residual (Dunn \& Smyth, 1996). It is a randomized version of the 
Cox \& Snell~(1968) residual and given by
\begin{equation}
\label{residalet}
r^q_t = \Phi^{-1}(u_t), \quad t=1,\ldots,n,
\end{equation} 
where $\Phi(\cdot)$ denotes the standard normal distribution function, $u_t$ is a uniform random variable on the interval $(a_t, b_t],$ with $a_t= \lim_{y\uparrow y_t} {\rm BI}_{c}(y_t;\widehat\alpha_t,\widehat\mu_t,\widehat\phi_t)$ and
$b_t = {\rm BI}_{c}(y_t;\widehat\alpha_t,\widehat\mu_t,\widehat\phi_t)$. Here,
${\rm BI}_{c}(y_t;\alpha_t,\mu_t,\phi_t)=\alpha_t {\rm 1\!l}_{[c,1]}(y_t)+(1-\alpha_t)F(y_t;\mu_t,\phi_t),$ where $F(\cdot;\mu_t,\phi_t)$ is the cumulative distribution function of the beta distribution ${\cal B}(\mu_t,\phi_t).$ 
In the zero-inflated beta regression model,
$u_t$ is a uniform random variable on $(0, \widehat\alpha_t]$ if $y_t=0$ and 
$u_t={\rm BI}_{0}(y_t;\widehat\alpha_t,\widehat\mu_t,\widehat\phi_t)$ if $y_t \in (0,1).$
On the other hand, in the one-inflated beta regression model, $u_t$ is a uniform random variable on $[\widehat\alpha_t,1)$ if $y_t=1$ and 
$u_t={\rm BI}_{1}(y_t;\widehat\alpha_t,\widehat\mu_t,\widehat\phi_t)$ if $y_t \in (0,1).$ Apart from sampling variability in $\widehat\alpha_t,$ $\widehat\mu_t,$ and $\widehat\phi_t$ the $r^q_t$ are  exactly standard normal in $(a_t, b_t]$ and the randomized procedure is introduced in order to produce a continuous residual. The randomized quantile residuals can vary from one  realization to another. In practice, it is useful to make at least four achievements. 

A plot of these residuals against the index of the observations $(t)$ should show no detectable pattern. A detectable trend in the plot of some residual against the predictors may be suggestive of link function misspecification. 
Also, normal probability plots with simulated envelopes are a helpful diagnostic tool (Atkinson, ~1985). 
Simulation results  not  presented here  indicated that the randomized quantile residuals perform well in detecting whether the distribution assumption is incorrect.

\paragraph{Global goodness-of-fit measure.} 
A simple global goodness-of-fit measure is a pseudo $R^2$, say $R_p^2$, defined by 
the square of the sample correlation coefficient between the outcomes, $y_1,\ldots,y_n$, 
and their corresponding predicted values, $\widehat{{\mu_1^\bullet}} , \ldots,\widehat{{\mu_n^\bullet}},$ where
$\widehat{{\mu^\bullet}_t}=\widehat{{\rm E}(y_t)}=c\widehat\alpha_t+(1-\widehat\alpha_t)\widehat\mu_t.$
A perfect agreement between the $y$'s and $\widehat\mu^\bullet$'s  yields $R_p^2=1$.
Other pseudo $R^2$'s are defined as $R^{2*}_p=1-\log \widehat L/\log \widehat L_0$ 
(McFadden,~1974) and $R^2_{{\rm LR}} = 1-(\widehat{L_0}/\widehat L)^{2/n}$ 
(Cox and Snell,~1989, p. 208-209), where $\widehat L_0$ and $\widehat L$ are the maximized 
likelihood functions of the null model and the fitted model, respectively.  
The ratio of the likelihoods or log-likelihoods may be regarded as measures of the improvement,
over the model with only three parameters ($\alpha$, $\mu,$ and $\phi$), achieved 
by the model under investigation.

\paragraph{Influence measures.}

A well-known measure of the influence of each observation on the regression parameter estimates is the likelihood displacement (Cook \& Weisberg, 1982, Ch.\ 3).
The likelihood displacement that results from removing the $t$th observation from the data is defined by 
$$
LD_t = \frac{2}{d}\{ \ell(\widehat\theta)-\ell(\widehat\theta_{(t)})  \},
$$
where $d$  is the dimension of $\theta$ and $\widehat\theta_{(t)}$ is the MLE of $\theta$ obtained after removing the $t$th observation from the data.
This definition does not consider that $\theta$ is actually split into two different types of parameters: the
parameters of the discrete component and the parameters of the continuous component.
Thus, it is more appropriate to consider the influence of the $t$th case on the estimation of $\rho$ and $(\beta, \gamma)$ separately. Therefore, we propose the following statistics: 
$$
\begin{aligned}
LD_t^{\rm D} &= \frac{2}{p}\{ \ell_1(\widehat\rho)-\ell_1(\widehat\rho_{(t)})  \}, \qquad t=0,1,\ldots,n \\
LD_t^{\rm C} &= \frac{2}{k+m}\{ \ell_2(\widehat\beta, \widehat\gamma)-\ell_2(\widehat\beta_{(t)}, \widehat\gamma_{(t)})  \}, \qquad t\!\!: y_t \in (0,1).
\end{aligned}
$$ 
Simple approximations for $LD_t^{\rm D}$ and $LD_t^{\rm C}$ are given by
$$
\begin{aligned}
LD_t^{\rm D} & \approx c_{tt}^{\rm D} = \frac{\mathbf{h}_{tt}}{p(1-\mathbf{h}_{tt})} (r_{pt}^{\rm D})^2,  \qquad t=0,1,\ldots,n \\
LD_t^{\rm C} & \approx c_{tt}^{\rm C} = \frac{\mathbf{P}_{tt}}{(k+m)(1-\mathbf{P}_{tt})} (r_{pt}^{\rm C})^2, \qquad t\!\!: y_t \in (0,1),  \\
\end{aligned}
$$
see Cook and Weisberg (1982, p. 191) and Wei (1998, p. 102). The approximations above are based on the iterative scheme for evaluating the MLE of
$\rho,$ $\beta,$ and $\gamma$ (Appendix \ref{AppB}) and Taylor expansions of $\ell_1(\widehat\rho_{(t)})$ and $\ell_2(\widehat\beta_{(t)}, \widehat\gamma_{(t)})$ around
$\widehat\rho$ and $(\widehat\beta, \widehat\gamma)$, respectively.
The quantities $c_{tt}^{\rm D}$ and $c_{tt}^{\rm C}$ can be useful to highlight influential cases.  In practice, we recommend the plotting of $ c_{tt}^{\rm D}$ and $ c_{tt}^{\rm C}$ against $t.$

\paragraph{Model selection.} 
Nested zero-or-one inflated beta regression models can be compared via the likelihood ratio test, using twice the difference between the maximized log-likelihoods of a full
model and a restricted model whose covariates are a subset of the full model. Information criteria, such as the generalized Akaike information criterion  (GAIC),  can be used for comparing non-nested models.  It is defined as 
\begin{equation}
\label{gaic}
{\rm GAIC} = \widehat{D}+ d \wp \ ,
\end{equation}
where $\widehat{D}=-2{\widehat\ell}$ is the global fitted deviance (Rigby \& Stasinopoulus, ~2005), $\widehat\ell$  is the maximized log-likelihood
and $d$ is the dimension of $\theta$. It is possible to interpret the first term on the right-hand side of (\ref{gaic}) as a measure of the lack of fit and the second term as a ``penalty" for adding $d$ parameters.
The model with the smallest GAIC is then selected. The Akaike information criterion AIC (Akaike, 1974), the Schwarz Bayesian criterion SBC (Schwarz, 1978), and the consistent
Akaike information criterion (CAIC) are special cases of GAIC corresponding to $\wp=2,$ $\wp=\log(n),$
and $\wp=\log(n)+1,$ respectively. 

\section{An application}
This section contains an application of the zero-inflated beta regression model to real data. We analyze data on the mortality in traffic accidents of 200 randomly selected Brazilian municipal
districts of the southeast region in the year 2002. The data were extracted from the DATASUS database available at \url{www.datasus.gov.bt}  and IPEADATA database available at \url{www.ipeadata.gov.br}.
The response variable $y$ is the proportion of deaths caused by traffic accidents. The explanatory variables are the logarithm of the number of inhabitants of the municipality ($lnpop$), the proportion
of the population living in the urban area ($propurb$), the proportion of men in the population ($propmen$), 
the  proportion of residents aged between 20 and 29 years ($prop2029$), and the human development
index of education of the municipal district $(hdie).$  The main objective is to investigate the effect of the 
young population ($prop2029$) on the proportion of deaths caused by traffic accidents
after controlling for potential confounding variables. We report the summary statistics on each of these 
variables in Table \ref{tabsummary}.

\begin{table}[!htb]
    \centering
    \caption{\label{tabsummary} Sample summary statistics.}

    \begin{center}
{\small
    \newcommand{\minitab}[2][l]{\begin{tabular}{#1}#2\end{tabular}}
    \renewcommand{\tabcolsep}{0.6pc} 
    \renewcommand{\arraystretch}{1.0} 
        \begin{tabular}{@{}rrrrrrrrr} \hline \hline

 variable & mean &  s.d. & min. & 1st qu. &  median &  3rd qu. & max.   \\ \hline\hline        
{\it lnpop}     &9.4042 &1.3464& 7.3920&8.3997&9.0620&10.0025& 16.1606  \\
{\it propurb}   &0.7173 &0.2044& 0.2042&0.5887&0.7445& 0.8862&  1.0000  \\ 
{\it propmen}   &0.5062 &0.0114& 0.4766&0.4989&0.5070& 0.5146&  0.5370 \\
{\it prop}2029  &0.1660 &0.0136& 0.1322&0.1559&0.1661& 0.1750&  0.2042  \\
{\it hdie}      &0.8236 &0.0587& 0.6060&0.7955&0.8325& 0.8640&  0.9330  \\ \hline\hline        

    \end{tabular}
}
    \end{center}

    \end{table}

\begin{figure}[!htb]
\begin{center}
\includegraphics[angle=270,scale=0.49]{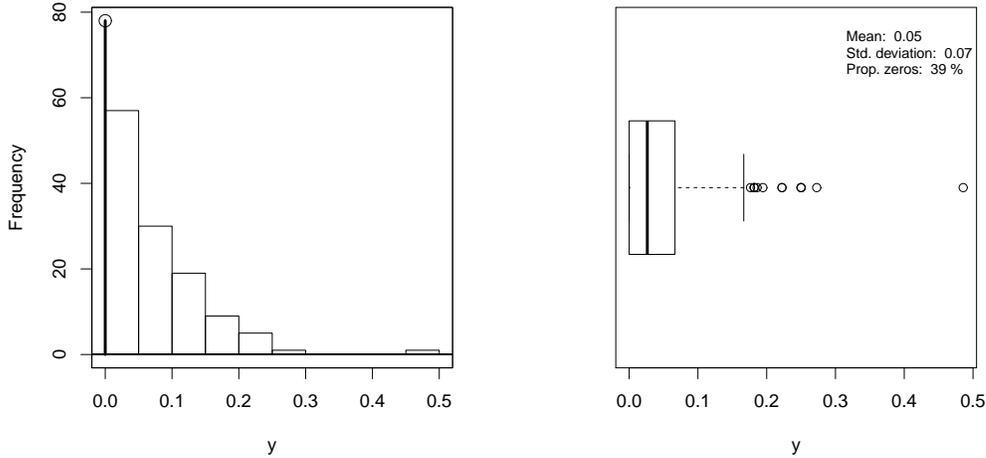}
\vspace{-0.5cm}
\caption{\label{descry} Frequency histogram and box-plot of the observed proportions of deaths caused by traffic accidents.}
\end{center}
\end{figure}

Figure \ref{descry} presents a histogram and a box-plot of the response variable. The clump-at-zero (the bar with the dot above) in the histogram represents 39\% of the data. We observe that the
distribution of the data in (0,1) is asymmetric with an inverted ``J" shape. Also, the box-plot reveals the presence of some outliers.  Visual inspection suggests that a zero-inflated beta
distribution may be a suitable model for the data.

We consider a zero-inflated beta regression model with the following specification: 
\begin{equation*}
            \label{modelosim1}
            \begin{aligned}
            {\rm logit}(\alpha) &=\rho_0+\rho_1\, lnpop+\rho_2\, propurb + \rho_3\, propmen + \rho_4\, prop2029 + \rho_5\, hdei, \\
            {\rm logit}(\mu)    &=\beta_0+\beta_1\, lnpop+\beta_2\, propurb + \beta_3\, propmen + \beta_4\, prop2029 + \beta_5\, hdei, \\
            \log(\phi)            &=\gamma_0+\gamma_1\, lnpop+\gamma_2\, propurb + \gamma_3\, propmen +\gamma_4\, prop2029 + \gamma_5\, hdei. \\
            \end{aligned}
\end{equation*}
Computations for fitting the model were carried out using the package {\tt gamlss} (Stasinopoulus \& Rigby,~2007) in
the {\tt R} software package (Ihaka \& Gentleman, 1996) and the {\tt BEZI} distribution implemented by Ospina (2006). The automatic procedure {\tt stepGAICAll.B()} included in the {\tt gamlss} package was used to perform model selection using the AIC.  The following final model has been selected:
\begin{equation}
            \label{modelosim2}
             \begin{aligned}
            {\rm logit}(\alpha) &=\rho_0+\rho_1\, lnpop+ \rho_4\, prop2029 + \rho_5\, hdei, \\
            {\rm logit}(\mu)    &=\beta_0+\beta_1\, lnpop+ \beta_4\, prop2029 + \beta_5\, hdei, \\
            \log(\phi)          &=\gamma_0+\gamma_1\, lnpop +\gamma_4\, prop2029 + \gamma_5\, hdei. \\
            \end{aligned}
\end{equation}
In fact, the value of the likelihood ratio statistic for testing 
$H_0\!\!: \theta=(\rho_2,\rho_3,\beta_2, \beta_3, \gamma_2, \gamma_3)=0$ versus a two-side alternative 
is $\Lambda =  4.89$ ($p$-value = $0.56$), i.e., $H_0$ is not rejected at the usual significance levels, and hence, {\it propurb} and {\it propmen} can be excluded from the model.

The parameter estimates and the corresponding standard errors for the final model are summarized in Table \ref{tabappli}.
The pseudo $R^2$'s are $R^{2*}_p= 0.597$  and $R^2_{{\rm LR}}= 0.584$, thus suggesting a reasonable fit of the model to the data. It is noteworthy that {\it prop2029}, which represents the proportion of the population aged between 20 to 29 years, 
has a positive effect on both the probability of observing at least one death due to traffic accidents ($1-\alpha$) and the 
conditional mean proportion of deaths due to this cause ($\mu$).
 
\begin{table}[!htb]
    \centering
    \caption{\label{tabappli} Parameter estimates (est.), standard errors (s.e.), and $p$-values ($p$).}
{\small
    \begin{center}
    \newcommand{\minitab}[2][l]{\begin{tabular}{#1}#2\end{tabular}}
    \renewcommand{\tabcolsep}{0.9pc} 
    \renewcommand{\arraystretch}{1.0} 

        \begin{tabular}{@{}crrc} \hline \hline
        
$\alpha$ & est. & s.e. & $p$ \\ \hline  
{\it intercept}& $   27.27  $ & $ 4.63  $ & $1.73{\rm e-}08 $ \\            
{\it lnpop}    & $   -1.17  $ & $ 0.27  $ & $2.07{\rm e-}05 $ \\            
{\it prop2029} & $   -48.06 $ & $ 17.46  $ & $6.49{\rm e-}03 $ \\            
{\it hdei}     & $  -11.34  $ & $ 4.01  $ & $5.13{\rm e-}03 $ \\\hline\hline      
$\mu$ & estimate & s.e. & $p$ \\ \hline                                                                         
{\it intercept}& $ -4.72    $ & $ 1.20  $ & $ 1.11{\rm e-}04$\\               
{\it lnpop}    & $ -0.53    $ & $ 0.05  $ & $ 7.14{\rm e-}17$\\               
{\it prop2029} & $  27.68   $ & $ 6.37  $ & $ 2.33{\rm e-}05$\\               
{\it hdei}     & $  3.10    $ & $ 1.44  $ & $ 3.28{\rm e-}02$ \\\hline\hline        
$\phi$ & estimate & s.e. & $p$ \\ \hline   
{\it intercept}& $  9.46   $ & $ 3.25   $ & $4.08{\rm e-}03$\\           
{\it lnpop}    & $  0.47   $ & $ 0.10   $ & $8.54{\rm e-}06$\\           
{\it prop2029} & $ -28.34  $ & $ 16.63  $ &  $9.01{\rm e-}02 $ \\         
{\it hdei}     & $ -6.70   $ & $ 3.90   $ &  $8.78{\rm e-}02 $ \\\hline\hline   

    \end{tabular}
    \end{center}
}
    \end{table}

\begin{figure}[!htb]
\begin{center}
\includegraphics[angle=270,scale=0.45]{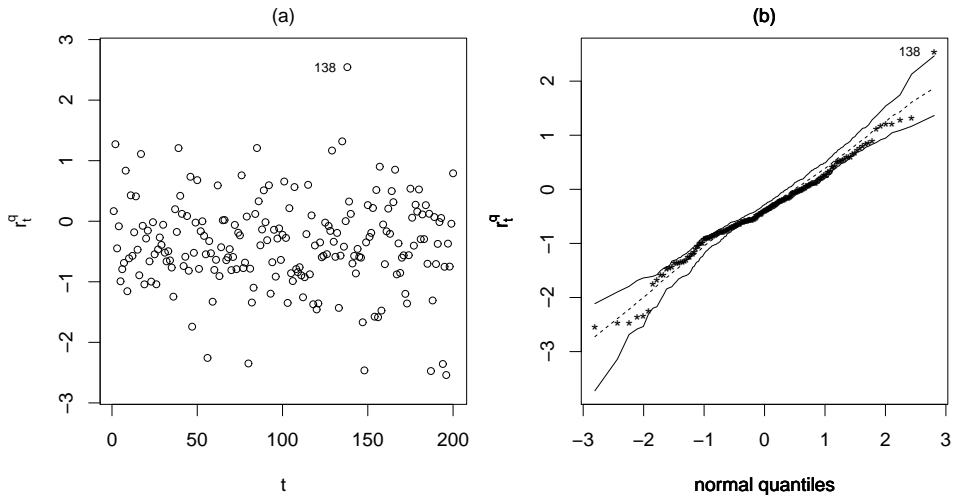}
\vspace{-0.5cm}
\caption{\label{residg} Quantile residual plots.}
\end{center}
\vspace{-0.3cm}
\end{figure}

Figure \ref{residg} shows the plot of randomized quantile residuals against the case number 
(Figure \ref{residg}(a)) and the quantile--quantile plot of the  ordered randomized quantile residuals against the 
corresponding quantiles of a standard normal distribution with a 95\% envelope based on 100 simulations 
(Figure \ref{residg}(b)). Figure \ref{residg}(a) singles out observation {\bf 138} as possibly atypical, and Figure \ref{residg}(b) indicates that it lies in the threshold of the envelope. 
Case 138 is, therefore, worthy of further investigation. Notice that Figure \ref{residg}(b) clearly shows that the 
distribution of the randomized quantile residuals is asymmetric, and the usual thresholds $(-2, \, 2)$ or $(-3, \, 3)$ 
should be used with care.

\begin{figure}[!htb]
\begin{center}
\includegraphics[angle=270,scale=0.56]{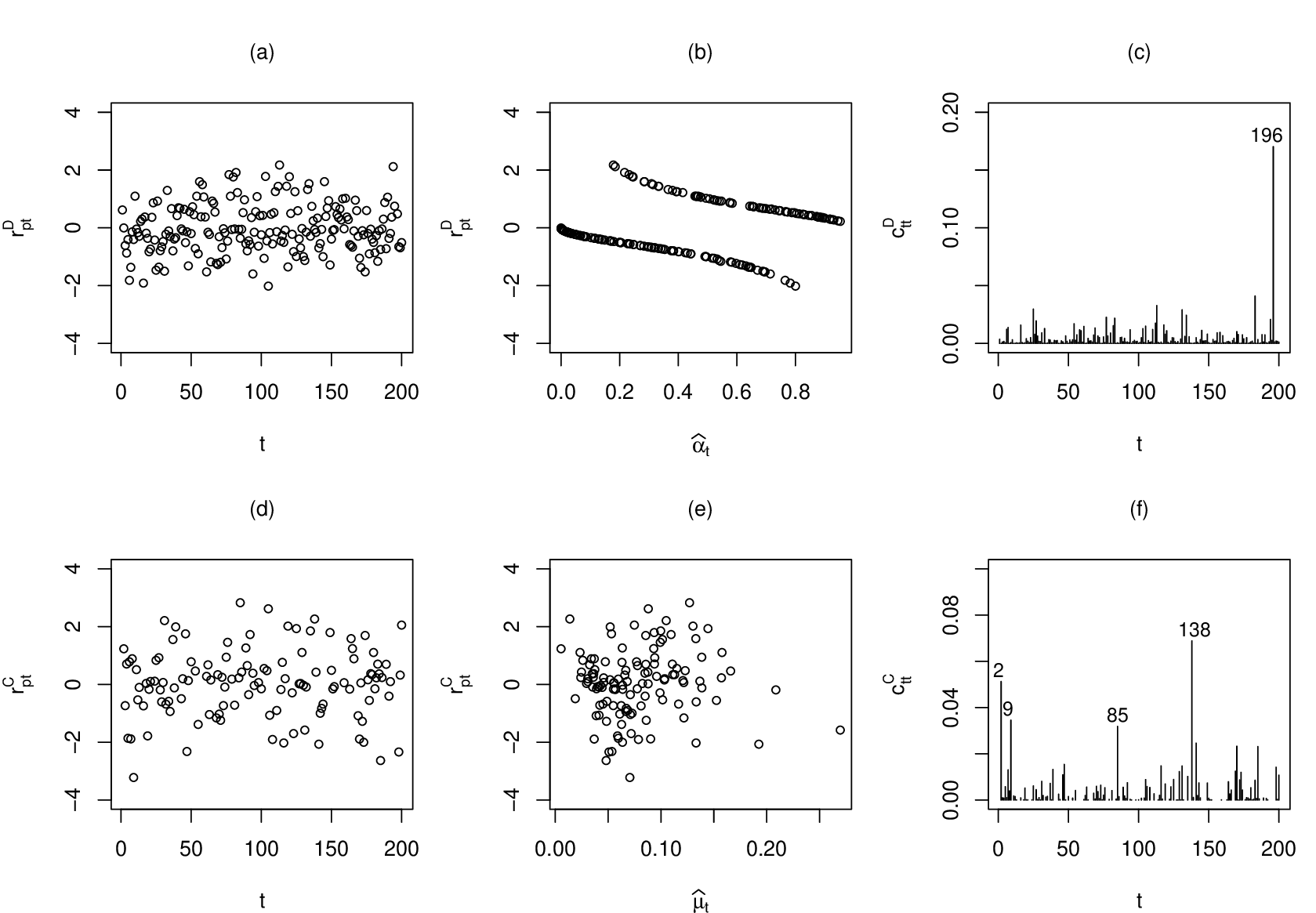}
\caption{\label{diagn} Diagnostic plots for the discrete component ((a)--(c)) and the continuous component ((d)--(f)).
}
\end{center}
\end{figure}

Diagnostic  plots for the discrete component are given in Figures \ref{diagn}(a)--\ref{diagn}(c).  The plot of $r_{pt}^{\rm D}$ against the case number (Figure \ref{diagn}(a)) suggests that the residuals appear to be randomly scattered around zero and 
that there is no atypical observation. The plot of $r_{pt}^{\rm D}$ against $\widehat\alpha_t$ (Figure \ref{diagn}(b)) is not suggestive of a lack of fit. The lower and upper bounds, corresponding to responses equal to zero and one, respectively, are typical of data with only two outcomes. Figure \ref{diagn}(c)  shows the plot of Cook statistics $c_{tt}^{\rm D}$ against the case number, from which observation {\bf 196} (an observation with the response variable equal to zero) is highlighted as influential.

Diagnostic  plots for the continuous component are given in Figures \ref{diagn}(d)--\ref{diagn}(f). Figure \ref{diagn}(d) shows the plot of $r_{pt}^{\rm D}$ against case number and suggests that the residuals are randomly scattered around zero. No  observation is distinguished as atypical. The plot of $r_{pt}^{\rm D}$ against $\widehat\mu_t$ (Figure \ref{diagn}(e))
is not suggestive of a lack of fit. Finally, the plot of  $c_{tt}^{\rm C}$ against the case number indicates  
observations {\bf 2}, {\bf 9}, {\bf85}, and {\bf 138} as influential points for the continuous component.

We have reestimated the model after removing the following sets of observations: \{138\}, \{196\}, and \{2, 9, 85, 138\}.
Our purpose is to investigate the impact of the exclusion of observations highlighted in the diagnostic plots 
on the inferences based on the estimated model. Table \ref{tabim} gives relative changes in the parameter estimates and the standard errors. Observation 138 alone is clearly  influential for the estimation of the continuous component of the model, while observation 196 solely impacts the inference on the discrete component, as expected. The joint exclusion of cases
2, 9, 85, and 138 produces substantial changes in the parameter estimates of the continuous component and only slight changes
in the estimated model for the discrete part. Our findings suggest that the proposed diagnostic tools are helpful for
detecting atypical observations and influential cases and are able to indicate the component of the model, discrete or continuous, that is affected by each influential observation.

\begin{table}[!h]
    \centering
    \caption{\label{tabim} Parameter estimates (est.), standard errors (s.e.), $p$-values ($p$), relative changes in estimates (rel. est.), and relative changes in standard errors (rel. s.e.)  because of exclusions of observations
    (in percentage).}
    \medskip
{\footnotesize
    \begin{center}
    \newcommand{\minitab}[2][l]{\begin{tabular}{#1}#2\end{tabular}}
    \renewcommand{\tabcolsep}{0.8pc} 
    \renewcommand{\arraystretch}{0.8} 

        \begin{tabular}{@{}ccrrcrr} \hline \hline
        
obs. & parameter & est. & s.e & $p$& rel. est. &  rel. s.e \\ \hline \hline    
    
                    &$\rho_0   $ & $  27.26$ & $  4.63$ & $ 0.00$ & $   0.04$ & $ -0.06$     \\      
                    &$\rho_1   $ & $  -1.17$ & $  0.27$ & $ 0.00$ & $   0.08$ & $ -0.17$     \\      
                    &$\rho_4   $ & $ -48.06$ & $ 17.46$ & $ 0.01$ & $  -0.01$ & $  0.01$     \\      
                    &$\rho_5   $ & $ -11.34$ & $  4.01$ & $ 0.01$ & $   0.04$ & $  0.00$     \\\cline{2-7}
                    &$\beta_0  $    & $  -5.05$ & $  1.11$ & $ 0.00$ & $  -6.96$ & $  7.06$   \\      
                    &$\beta_1  $    & $  -0.57$ & $  0.06$ & $ 0.00$ & $  -6.85$ & $ -1.48$   \\      
138                 &$\beta_4  $    & $  32.40$ & $  6.42$ & $ 0.00$ & $ -17.07$ & $ -0.67$   \\      
                    &$\beta_5  $    & $   2.93$ & $  1.40$ & $ 0.04$ & $   5.54$ & $  2.64$   \\\cline{2-7}
                    &$\gamma_0 $    & $  10.66$ & $  3.00$ & $ 0.00$ & $ -12.60$ & $  7.74$   \\      
                    &$\gamma_1 $    & $   0.61$ & $  0.11$ & $ 0.00$ & $ -27.20$ & $ -7.02$   \\      
                    &$\gamma_4 $    & $ -43.46$ & $ 17.26$ & $ 0.01$ & $ -53.34$ & $ -3.78$   \\      
                    &$\gamma_5 $    & $  -6.47$ & $  3.83$ & $ 0.09$ & $   3.52$ & $  2.02$   \\\hline \hline 
                                   
                    &$\rho_0   $ & $  30.78$ & $  5.25$ & $ 0.00$ & $ -12.86$ & $ -13.34    $  \\      
                    &$\rho_1   $ & $  -1.38$ & $  0.30$ & $ 0.00$ & $ -18.15$ & $ -12.33    $  \\      
                    &$\rho_4   $ & $ -51.62$ & $ 18.21$ & $ 0.01$ & $  -7.41$ & $  -4.30    $  \\      
                    &$\rho_5   $ & $ -12.68$ & $  4.24$ & $ 0.00$ & $ -11.80$ & $  -5.95    $  \\\cline{2-7}
                    &$\beta_0  $    & $  -4.72$ & $  1.20$ & $ 0.00$ & $   0.00$ & $   0.00 $  \\      
                    &$\beta_1  $    & $  -0.53$ & $  0.06$ & $ 0.00$ & $   0.00$ & $   0.00 $  \\      
196                 &$\beta_4  $    & $  27.68$ & $  6.38$ & $ 0.00$ & $   0.00$ & $   0.00 $  \\      
                    &$\beta_5  $    & $   3.10$ & $  1.44$ & $ 0.03$ & $   0.00$ & $   0.00 $  \\\cline{2-7}
                    &$\gamma_0 $    & $   9.47$ & $  3.26$ & $ 0.00$ & $   0.00$ & $   0.00 $  \\      
                    &$\gamma_1 $    & $   0.48$ & $  0.10$ & $ 0.00$ & $   0.00$ & $   0.00 $  \\      
                    &$\gamma_4 $    & $ -28.34$ & $ 16.64$ & $ 0.09$ & $   0.00$ & $   0.00 $  \\      
                    &$\gamma_5 $    & $  -6.70$ & $  3.91$ & $ 0.09$ & $   0.00$ & $   0.00 $  \\\hline \hline 
 
                    &$\rho_0   $ & $   26.97$ & $  4.62$ & $0.00$ & $ 1.11$ & $ 0.17 $     \\      
                    &$\rho_1   $ & $   -1.16$ & $  0.27$ & $0.00$ & $ 1.17$ & $ 0.24 $     \\      
                    &$\rho_4   $ & $  -47.28$ & $ 17.45$ & $0.01$ & $ 1.63$ & $ 0.06 $     \\      
                    &$\rho_5   $ & $  -11.27$ & $  4.00$ & $0.01$ & $ 0.67$ & $ 0.09 $     \\\cline{2-7}
                    &$\beta_0  $    & $  -3.98$ & $   1.11$ & $0.00$ & $ 15.69$ & $  6.93 $   \\      
2, 9, 85, 138       &$\beta_1  $    & $  -0.58$ & $   0.05$ & $0.00$ & $ -8.90$ & $  5.45 $   \\      
                    &$\beta_4  $    & $  29.71$ & $   5.87$ & $0.00$ & $ -7.33$ & $  8.03 $   \\      
                    &$\beta_5  $    & $   2.29$ & $   1.36$ & $0.09$ & $ 26.14$ & $  5.94 $   \\\cline{2-7}
                    &$\gamma_0 $    & $   7.48$ & $   3.35$ & $0.03$ & $ 20.96$ & $ -2.82 $   \\      
                    &$\gamma_1 $    & $   0.61$ & $   0.14$ & $0.00$ & $-27.75$ & $-36.20 $   \\      
                    &$\gamma_4 $    & $ -29.35$ & $  16.95$ & $0.08$ & $ -3.57$ & $ -1.86 $   \\       
                    &$\gamma_5 $    & $  -5.37$ & $   4.17$ & $0.20$ & $ 19.85$ & $ -6.70 $   \\\hline \hline 

      \end{tabular}     
    \end{center}      
}                     
    \end{table}

\section{Concluding remarks}

We developed a general class of zero-or-one inflated beta regression models that can be useful for practitioners when modeling response variables in the standard unit interval, such as rates or proportions, with the presence of zeros or ones. Explicit formulas for the score function, Fisher's information matrix, and its inverse are given.  An iterative estimation procedure and its computational implementation are discussed. Interval estimation for different population quantities  is presented. We also proposed a set of diagnostic tools that can be employed to identify departures from the
postulated model and the atypical and influential observations. These tools include pseudo $R^2$'s, different residuals, influence measures, and model selection procedures. One particularly interesting feature of the proposed diagnostic analysis is that it allows the practitioner to separately identify influential cases on the discrete 
and the continuous components of the model. An application using real data was presented and discussed.

\section*{Acknowledgements} We gratefully acknowledge partial financial support from FAPESP/Brazil and CNPq/Brazil. 

\vspace{0.6cm}
\appendix
{\small
\section{Appendix A: Score vector and observed information matrix}
\paragraph{Score vector.} The elements of the score vector are given by
\begin{equation}
\label{Us}
U_s = 
{{\partial\ell(\theta)}\over{\partial\rho_s}} = 
\SumD\frac{\partial\ell_t(\alpha_t)}{\partial\alpha_t}
     \frac{\partial\alpha_t}{\partial\eta_{1t}}
     \frac{\partial\eta_{1t}}{\partial\rho_s}
=\sum_{t=1}^n \frac{{{\rm 1\!l}_{\{c\}}(y_t)}-\alpha_t}{\alpha_t(1-\alpha_t)}\frac{1}{h'_1(\alpha_t)} \frac{\partial \eta_{1t}}{\partial \rho_s},
\end{equation}
\begin{equation}
\label{Ur}
\begin{aligned}
U_r &= 
{{\partial\ell(\theta)}\over{\partial\beta_r}} = 
\SumA \
\frac{\partial\ell_t(\mu_t, \phi_t)}{\partial\mu_t}
\frac{\partial\mu_t}{\partial\eta_{2t}}
\frac{\partial\eta_{2t}}{\partial\beta_r} =
\SumA \
 \phi_t(y_t^* - \mu^*_t)
\frac{1}{h'_2(\mu_t)} \frac{\partial \eta_{2t}}{\partial \beta_r},
\end{aligned}
\end{equation} 
\begin{equation}
\label{UR}
\begin{aligned}
U_R &= 
{{\partial\ell(\theta)}\over{\partial\gamma_R}} = 
\SumA \
\frac{\partial\ell_t(\mu_t, \phi_t)}{\partial\phi_t}
\frac{\partial\phi_t}{\partial\eta_{3t}}
\frac{\partial\eta_{3t}}{\partial\gamma_R} =
\SumA \
[\mu_t(y_t^* - \mu^*_t)+(y_t^\dagger - \mu^\dagger_t)]
\frac{1}{h'_3(\phi_t)} \frac{\partial \eta_{3t}}{\partial \gamma_R},
\end{aligned}
\end{equation} 
for $s=1,\ldots,p$; $r=1,\ldots, k$; and $R=1,\ldots, m.$

\paragraph{Observed information matrix.}
For $s, s'=1,\ldots,p$; $r,r'=1,\ldots, k;$ and $R,R'=1,\ldots, m$, we have 
\begin{equation*}
\label{Us_s}
\begin{aligned}
 J_{ss'} &= -{{\partial^2\ell(\theta)}\over{\partial\rho_s\partial\rho_{s'}}} =
-\sum_{t=1}^n 
\Bigg\{
    \Bigg(
        \frac{-{{\rm 1\!l}_{(0,1)}(y_t)}}{(1-\alpha_t)^2}
        -\frac{{{\rm 1\!l}_{\{c\}}(y_t)}}{\alpha_t^2}
    \Bigg)\Bigg[\frac{1}{h'_1(\alpha_t)}\Bigg]^2 \\
    & + \Bigg(\frac{{{\rm 1\!l}_{\{c\}}(y_t)}-\alpha_t}{\alpha_t(1-\alpha_t)}
      \Bigg)
\Bigg[\frac{1}{h'_1(\alpha_t)}\Bigg] \Bigg[\frac{-h''_1(\alpha_t)}{(h'_1(\alpha_t))^2}\Bigg]
\Bigg\} \frac{\partial\eta_{1t}}{\partial\rho_{s}}\frac{\partial\eta_{1t}}{\partial\rho_{s'}} \\
&-\sum_{t=1}^n 
\Bigg\{
\frac{{{\rm 1\!l}_{\{c\}}(y_t)}-\alpha_t}{\alpha_t(1-\alpha_t)}\Bigg[\frac{1}{h'_1(\alpha_t)}\Bigg]\frac{\partial^2\eta_{1t}}{\partial\rho_{s}\partial\rho_{s'}}\Bigg\},
\end{aligned}
\end{equation*} 
\begin{equation*}
\label{Ur_r}
\begin{aligned}
J_{rr'} = 
-\frac{\partial^2\ell(\theta)}{\partial\beta_r\partial\beta_{r'}}&=
- \!\!\!\!\!\! \SumA \ 
\Bigg\{
-\phi_t^2v_t^*\frac{1}{h'_2(\mu_t)} +\phi_t(y_t^* - \mu^*_t)\Bigg(\frac{-h''_2(\mu_t)}{(h'_2(\mu_t))^2}\Bigg)
\frac{1}{h'_2(\mu_t)}\frac{\partial\eta_{2t}}{\partial\beta_{r}}\frac{\partial\eta_{2t}}{\partial\beta_{r'}}
\Bigg\}  \\
&- \!\!\!\!\!\!
\SumA \ 
\Bigg\{
\phi_t(y_t^* - \mu^*_t)\frac{1}{(h'_2(\mu_t))}
\frac{\partial^2\eta_{2t}}{\partial\beta_{r}\partial\beta_{r'}}\Bigg\},
\end{aligned}
\end{equation*}

\begin{equation*}
\label{Ur_R}
J_{rR}= 
-\frac{\partial^2\ell(\theta)}{\partial\beta_r\partial\gamma_{R}}=
-\!\!\!\!\!\! \SumA \ 
\Bigg\{
[y_t^* - \mu^*_t  -\phi_t(\mu_t v_t^*+c_t^{\ast\dagger})] 
\frac{1}{(h'_2(\mu_t))}\frac{\partial\eta_{2t}}{\partial\beta_{r}}
\frac{1}{(h'_3(\phi_t))}\frac{\partial\eta_{3t}}{\partial\gamma_{R}}
\Bigg\},
\end{equation*} 
\begin{equation*}
\label{UR_R}
\begin{aligned}
J_{RR'} = -\frac{\partial^2\ell(\theta)}{\partial\gamma_{R}\partial\gamma_{R'}}&= 
-\!\!\!\!\!\!\SumA \ 
\Bigg\{
[-\mu_t^2 v_t^* -2\mu_t c^{\ast\dagger} - v_ t^\dagger]\frac{1}{(h'_3(\phi_t))}
+ [\mu_t(y_t^* - \mu^*_t)+(y_t^\dagger - \mu^\dagger_t)]\\
&\times \Bigg(\frac{-h''_3(\phi_t)}{(h'_3(\phi_t))^2}\Bigg)
\Bigg\}\frac{1}{(h'_3(\phi_t))} \frac{\partial\eta_{3t}}{\partial\gamma_{R}} \frac{\partial\eta_{3t}}{\partial\gamma_{R'}}
\\
&- \!\!\!\!\!\!
\SumA \ 
\Bigg\{
[\mu_t(y_t^* - \mu^*_t)+(y_t^\dagger - \mu^\dagger_t)] \frac{1}{(h'_3(\phi_t))} 
\frac{\partial^2\eta_{3t}}{\partial\gamma_{R}\partial\gamma_{R'}}\Bigg\}. 
\end{aligned}
\end{equation*}
The observed Fisher information matrix can now be written as  
\begin{equation}\label{J}
 J(\theta) = 
\begin{pmatrix}
J_{\rho\rho} & 0 & 0 \\
0            &  J_{\beta\beta} &  J_{\beta\gamma} \\
0            &  J_{\gamma\beta} &  J_{\gamma\gamma} \\
\end{pmatrix},
\end{equation}
where
\begin{equation*}
\label{Jrr}
J_{\rho\rho}     =  {\cal V}^\top \{ ({\cal A}^{*2}(I_n - Y^c)+{\cal A}^2 Y^c) + {\cal A}^{*}{\cal A}(Y^c -\alpha^*){\cal D}^*{\cal D}  \}{\cal V}^\top - [ (y^c -\alpha)^\top {\cal A}{\cal D} ][\dot{\cal V}],\\
\end{equation*} 
\begin{equation*}
\label{Jss}
 J_{\beta\beta}   = {\cal X}^\top \{\Phi T V^* +S T^2(Y^* - {\cal M}^* )\}(I_n-Y^c)T \Phi  {\cal X} +[(y^* -\mu^*)^\top (I_n-Y^c)T \Phi ][\dot{\cal X}],
\end{equation*} 
\begin{equation*}
\label{Jrs}
J_{\gamma\beta} = {J^\top_{\gamma\beta}} = -{\cal X}^\top \{(Y^* - {\cal M}^* ) - \Phi({\cal M} V^* + C )\} (I_n-Y^c) T H {\cal Z},
\end{equation*} 
\begin{equation*}
\label{JRR}
\begin{aligned}
J_{\gamma\gamma} &=  
{\cal Z}^\top \{ H ({\cal M}^2  V^* +2{\cal M} C + V^\dagger) 
+\{ {\cal M} (Y^* - {\cal M}^*) + (Y^\dagger - {\cal M}^\dagger)\} Q H^2\} (I_n-Y^c) H {\cal Z} \\
& +[((y^* - \mu^*)^\top {\cal M} \, + \,(y^\dagger - \mu^\dagger)^\top)(I_n-Y^c) H ][\dot{\cal Z}],
 \end{aligned}
\end{equation*} 
with 
$\alpha^* = {\rm diag}(\alpha_1, \ldots, \alpha_n),$ 
$Y^*=  {\rm diag}(y^*_1, \dots, y^*_n),$  
$Y^\dagger=  {\rm diag}(y^\dagger_1, \dots, y^\dagger_t),$ 
${\cal M}^* = {\rm diag}(\mu_1^*, \ldots, \mu_n^*),$ 
${\cal M}^\dagger = {\rm diag}(\mu_1^\dagger, \ldots, \mu_n^\dagger),$ and 
$Q =  {\rm diag}(h_3''(\phi_1), \ldots,h_3''(\phi_n))$;  $\dot{\cal V}$ is an $n\times p \times p$ array with faces 
${\cal V}_t = {\partial^2\eta_{1t}}/{\partial\rho\partial\rho^\top},$ $\dot{\cal X}$ is an $n\times k \times k$ array with faces 
${\cal X}_t = {\partial^2\eta_{2t}}/{\partial\beta\partial\beta^\top},$ and  $\dot{\cal Z}$ is an $n\times m \times m$ array with faces 
${\cal Z}_t = {\partial^2\eta_{3t}}/{\partial\gamma\partial\gamma^\top},$ for $t=1,\ldots,n.$ The column multiplication for three-dimensional arrays is indicated by the bracket operator $[\cdot][\cdot]$ as defined
by Wei~(1998, p.\ 188).
}

\vspace{0.6cm}
{\small
\section{Appendix B: Iterative algorithm for maximum likelihood estimation}
\label{AppB}

MLEs for $\rho$ and $\vartheta=(\beta^\top, \phi)^\top$ can be obtained by using a re-weighted least-squares algorithm. For $\rho$ we have 
\begin{equation}
\label{pros1}
\rho^{(m+1)} =  ({\cal V}^{(m)\top }\mathbf{{W_1}}^{(m)}{\cal V}^{(m)})^{-1}{\cal V}^{(m)}\mathbf{{W_1}}^{(m)}\mathbf{{y_1}}^{(m)}, \qquad m=0,1,\ldots,
\end{equation}
where 
$\mathbf{{W_1}}$ is defined in \eqref{fisher} and
\begin{equation}
\label{reglocal}
\mathbf{{y_1}} = {\cal V}\rho + \mathbf{{W_1}}^{-1}{\cal A} { D} {\cal A}^*(y^c - \alpha)
\end{equation}
is a local modified dependent variable. This cycle is repeated until convergence is achieved. Rewriting this equation as
\begin{equation}
\mathbf{{y_1}} = \eta_1 - \tau_1 +  \mathbf{{W_1}}^{-1}{\cal A} { D} {\cal A}^*(y^c - \alpha),
\end{equation}
where $\eta_1 = \textbf{\textit{f}}_{\bf 1}( V;\rho)$ and $\tau_1 = \textbf{\textit{f}}_{\bf 1}( V;\rho)-{\cal V}\rho,$ we can interpret \eqref{pros1} as an iterative process to fit a generalized linear model with design matrix ${\cal V},$ 
systematic part $h_1(\alpha_t)=\eta_{1t},$ $t$th diagonal element of the variance function $[\mathbf{{W_1}}^{-1}{\cal A} { D} {\cal A}^*]_{tt},$ $t=1, \ldots, n,$ and  offset $\tau_1.$ The offset quantity is subtracted, at each step, from the predictor $\eta_1.$   The iterative process \eqref{pros1} may be performed, for instance, in the {\tt R} package by taking advantage of the library {\tt MASS} (Venables \& Ripley, 2002). 
This procedure allows us to extend the diagnostic results of ordinary regression models to the discrete component of the zero-or-one inflated beta regression model. Upon the converge of iterative process \eqref{pros1}, we have
$
\widehat\rho =  ({\widehat{\cal V}}^\top \mathbf{{\widehat{W}_1}} \widehat{\cal V} )^{-1} \widehat{\cal V}^\top \mathbf{z},
$
where
$
\mathbf{z} = \widehat{\cal V}\widehat\rho \, + \, {\mathbf{{\widehat{W}_1}}^{-1}} \widehat{\cal A} \widehat{ D} \widehat{\cal A}^*(y^c - \widehat\alpha).
$ The ordinary residual for this re-weighted regression is 
\begin{equation*}
r^*= {\mathbf{{\widehat{W}_1}}^{1/2}}(\mathbf{z} -\widehat\eta_1) = {\mathbf{{\widehat{W}_1}}^{-1/2}} \widehat{\cal A} \widehat{ D} \widehat{\cal A}^*(y^c - \widehat\alpha).
\end{equation*}
Note that the $t$th element of $r^*$ is 
\begin{equation}
\label{rescd}
r_t^* = \frac{ {\rm 1\!l}_{\{c\}} - \widehat\alpha_t }{ \sqrt{\widehat\alpha_t(1-\widehat\alpha_t)} }.
\end{equation} 
By writing 
$({\widehat{\cal V}}^\top \mathbf{{\widehat{W}_1}} \widehat{\cal V} )\widehat\rho = \widehat{\cal V}^\top \mathbf{z},$ it is possible to obtain the approximations ${\rm E}(r^*)\approx 0$ and ${\rm Var}(r^*)\approx(I_n- {\mathbf{{H}}}),$ where 
$I_n$ is the $n \times n$ identity matrix and 
\begin{equation}
\label{hm}
\mathbf{{H}}=\mathbf{{W_1}}^{1/2} {\cal V} ( {\cal V}^\top \mathbf{{W_1}} {\cal V} )^{-1} {\cal V}^\top \mathbf{{W_1}}^{1/2}
\end{equation}
is an orthogonal projection matrix onto the vector space spanned by the columns of ${\cal V}.$ The geometric interpretation of $\mathbf{{H}}$ as a projection matrix is discussed by Moolgavark {\it et al}.~(1984).  

Now, let $\Xtil,$ $\Ttil,$ and $\Wtil$ be the $(2n \times (k+n))$ and $(2n \times 2n),$ $(2n \times 2n)$ 
dimensional matrices
\begin{equation}
\Xtil =
\begin{pmatrix}
 {\cal X} &  0 \\
0 &  {\cal Z}\\
 \end{pmatrix}, \qquad 
\Ttil =
\begin{pmatrix}
(I_n-Y^c) T\Phi &  0 \\
0 &  (I_n-Y^c) H\\
 \end{pmatrix}, \qquad 
\Wtil =
\begin{pmatrix}
 \mathbf{{W_2}} & \mathbf{{W_3}} \\
\mathbf{{W_3}} &  \mathbf{{W_4}}\\
 \end{pmatrix}, 
\end{equation}
respectively, and $\ytil^\top = ((y^*-\mu^*)^\top,[{\cal M}(y^*-\mu^*)+(y^\dagger-\mu^\dagger)]^\top)$ be an $1 \times 2n$ auxiliary vector. The score vector corresponding to $\vartheta =(\beta^\top, \gamma^\top)^\top$ can be written as 
\begin{equation}
 \label{escvar:theta}
U(\vartheta) = \, \,\Xtil^\top \Ttil \ytil.
\end{equation}
Fisher's information matrix for the parameter vector $\vartheta$ is given by  
\begin{equation}
K(\vartheta) = \,\,\Xtil^\top \Wtil \Xtil.
\end{equation}
Also, the iterative process for estimating $\vartheta$  takes the form  
\begin{equation}
\label{pros2}
\vartheta^{(m+1)} =  (\Xtil^{(m)\top }\Wtil^{(m)}\Xtil^{(m)})^{-1}\Xtil^{(m)}\Wtil^{(m)}\ytil^{*(m)}, 
\end{equation}
where $\ytil^{*(m)} \,\,= \,\,\Xtil \vartheta \,\,+ \Wtil^{-1}\Ttil \ytil$ 
and  $m = 0, 1, 2,\ldots$ are the iterations that are performed until convergence, which occurs when the distance between $\vartheta^{(m+1)}$ and $\vartheta^{(m)}$ becomes smaller than a given, small constant. The procedure is initialized  by choosing suitable initial values for $\beta$ and $\gamma.$

Assuming that $\gamma$ is known, Fisher's scoring iterative scheme used to estimating $\beta$ can be written as \begin{equation}
\label{betam}
\beta^{(m+1)} =  ({\cal X}^{(m)\top }\mathbf{{W_2}}^{(m)}{\cal X}^{(m)})^{-1}{\cal X}^{(m)}\mathbf{{W_2}}^{(m)}\mathbf{{y_2}}^{(m)}, 
\end{equation}
where $\mathbf{{W_2}}$ is defined in \eqref{fisher} and $\mathbf{{y_2}} = {\cal X}\beta + \mathbf{{W_2}}^{-1}T \Phi (y^*-\mu^*),$ with $m = 0, 1, 2,\ldots.$ Upon convergence, 
we have
\begin{equation}
\label{betacon}
\widehat\beta =  ({\widehat{\cal X}}^\top \mathbf{{\widehat{W}_2}} \widehat{\cal X} )^{-1} \widehat{\cal X}^\top \mathbf{\tau},
\end{equation}
where
$
\mathbf{\tau} = \widehat{\cal X}\widehat\beta \,+\, {\mathbf{{\widehat{W}_2}}^{-1}} \widehat T \widehat{ \Phi}(y^*-\widehat\mu^*).
$  Then, $\widehat\beta$ in \eqref{betacon} can be viewed as the least-squares estimate of $\beta$ obtained by regressing $\mathbf{\tau}$ on ${\cal X}$ with weighting matrix $\mathbf{W_2}$. The ordinary residual of this re-weighted least-squares regression is 
\begin{equation}
\label{resid}
r = (I_n- {\mathbf{{\widehat P}}})\tau = {\mathbf{{\widehat W}_2}}^{1/2}(\tau -{\widehat{\cal X}}\widehat\beta  )={\mathbf{{{\widehat W}_2}}}^{-{1/2}}\widehat\Phi \widehat T (y^*-\widehat\mu^*). 
\end{equation}
Here, 
\begin{equation}
\label{hatP}
{\mathbf{{\widehat P}}}=
{\mathbf{{{\widehat W}_2}}}^{{1/2}}   {{\widehat\cal X}} ({\widehat{\cal X}}^\top {\mathbf{{{\widehat W}_2}}} {\widehat{\cal X}} )^{-1} {\widehat{\cal X}}^\top  {\mathbf{\widehat W}_2}^{{1/2}}  
\end{equation}
 is a projection matrix. Note that if all the quantities are evaluated at the true parameter, ${\rm E}(r)=0$ and ${\rm Var}(r)=(I_n- {\mathbf{{ P}}}).$ The ${\mathbf{{ P}}}$ matrix is similar to the leverage matrix in standard linear regression models, and hence, we refer to it as the
generalized leverage matrix. It is possible to show that $I_n-{\mathbf{{ P}}}$ is symmetric,
idempotent, and spans the residual $r$-space. This implies that a small $1-{\mathbf{{ P}}_{tt}}$ indicates
extreme points in the design space of the continuous component of the zero-or-one inflated beta regression model.
Note that the $t$th element of $r$ is 
\begin{equation}
\label{r}
r_t= \frac{y^*-\widehat\mu^*}{\sqrt{\widehat{v}^*_{t}  (1-\widehat\alpha_t)}}.
\end{equation}
}

\section*{References}
\begin{description}
\small

\R{Akaike, H.}~(1974). A new look at the statistical model identification.
{\it IEEE. Transactions on Automatic Control,}  {\bf 19}, 716--723.

\R{Atkinson, A. C.}~(1985). {\it Plots, Transformations and Regression: An Introduction to Graphical Methods of
Diagnostic Regression Analysis}. New York: Oxford University Press.



\R{Cook, R. D. \&  Weisberg, S.}~(1982). {\it Residuals and Influence in Regression.} London: Chapman and Hall. 
 
\R{Cook, D. O., Kieschnick, R. \& McCullough, B. D.}~(2008). Regression analysis of proportions in finance with self selection.
{\it Journal of Empirical Finance}, {\bf 15}, 860--867.


\R{Cox, D. R. \& Hinkley, D. V.}~(1974). {\it Theoretical Statistics}. London: Chapman and Hall.

\R{Cox, D. R. \& Reid, N.}~(1987). Parameter orthogonality and approximate conditional inference (with discussion).
{\it Journal of the Royal Statistical Society} {\bf B}, {\bf 49}, 1--39.

\R{Cox, D. \& Snell, E.}~(1968). A general definition of residuals. {\it Journal of the Royal Statistical Society} {\bf B},
{\bf 30}, 248--275.

\R{Cox, D. R. \& Snell, E. J.}~(1989). {\it Analysis of Binary Data}. London: Chapman and Hall.



\R{Dunn, P. K.  \& Smyth, G. K.}~(1996). Randomized quantile residuals. {\it Journal of Computational and Graphical Statistics},
{\bf 5}, 236--244.


\R{Espinheira, P. L., Ferrari, S. L. P. \& Cribari--Neto, F.}~(2008a). 
Influence diagnostics in beta regression. 
{\it Computational Statistics and Data Analysis}, {\bf 52}, 4417--4431. 

\R{Espinheira, P. L., Ferrari, S. L. P. \& Cribari--Neto, F.}~(2008b).
On beta regression residuals. 
{\it Journal of Applied Statistics}, {\bf 35}, 407--419. 

\R{Fahrmeir, L. \& Kaufmann, H.}~(1985). Consistency and asymptotic normality of the maximum likelihood estimator in generalized linear models. {\it Annals of Statistics}, {\bf 13}, 342--368. 

\R{Ferrari, S. L. P. \& Cribari--Neto, F.}~(2004). Beta regression for modelling rates and proportions.
 {\it Journal of Applied Statistics,} 
{\bf 7}, 799--815.

\R{Ferrari, S. L. P. \& Pinheiro, E. C.}~(2010). Improved likelihood inference in beta regression.
{\it Journal of Statistical Computation and Simulation}. Available online. {\tt DOI: 10.1080/00949650903389993}.

\R{Hoff, A.}~(2007). Second stage DEA: Comparison of approaches for modelling the DEA score.
{\it European Journal of Operational Research,} {\bf 181}, 425--435.

\R{Ihaka, R. \& Gentleman, R.}~(1996).  R: A language for data analysis and graphics. 
{\it Journal of Computational and Graphical Statistics}, {\bf 5,} 299--314.

\R{Johnson, N., Kotz, S. \& Balakrishnan, N.}~(1995). 
{\it Continuous Univariate Distributions}. 2nd  ed.
New York: John Wiley and Sons. 

\R{Kieschnick, R. \& McCullough, B. D.}~(2003). Regression analysis of variates
observed on (0,1): percentages, proportions, and fractions.
{\it Statistical Modelling,} {\bf 3}, 1--21. 

\R{Korhonen, L., Korhonen, K. T., Stenberg, P., Maltamo, M. \& Rautiainen, M.}~(2007).
Local models for forest canopy cover with beta regression.
{\it Silva Fennica}, {\bf 41}, 671--685. 

\R{McCullagh, P. \& Nelder, J. A.}~(1989). {\it Generalized Linear Models},
2nd ed. London: Chapman and Hall.


\R{McFadden, D.}~(1974). Conditional logit analysis of qualitative choice behavior. In: P. Zarembka,
ed., {\it Frontiers in Econometrics}, 105--142. New York: Academic Press.


\R{Ospina, R.}~(2006). The zero-inflated beta distribution for fitting a GAMLSS.  Extra distributions to be used for GAMLSS modelling. Available at {\tt gamlss.dist} package. {\tt http://www.gamlss.org}.

\R{Ospina, R. \& Ferrari, S. L. P.}~(2010). Inflated beta distributions. {\it Statistical Papers},
{\bf 51}, 111--126. 


\R{Pace, L. \& Salvan, A.}~(1997). {\it Principles of Statistical Inference from a Neo-Fisherian Perspective.} Advanced Series on Statistical Science and Applied Probability, Vol.4. Singapore: World Scientific. 

\R{Paolino, P.}~(2001). Maximum likelihood estimation of models with beta-distributed 
dependent variables. {\it Political Analysis}, {\bf 9}, 325--346. 


\R{Press, W. H., Teulosky, S. A., Vetterling, W. T. \& Flannery, B. P.}~(1992).
{\it Numerical Recipes in C: The Art of Scientific Computing}. 2nd  ed.
Prentice Hall: London.

\R{Rao, C. R.}~(1973). {\it Linear Statistical Inference and Its Applications}, 2nd ed. New York: Wiley.

\R{Rigby, R. A. \& Stasinopoulos, D. M.}~(2005). Generalized additive models for location, scale and shape 
(with discussion). {\it Applied Statistics}, {\bf 54}, 507--554.

\R{Schwarz, G.}~(1978). Estimating the dimension of a mode. {\it Annals of Statistics}, {\bf 6}, 461--464.

\R{Simas, A. B., Barreto-Souza, W. \& Rocha, A. V.}~(2010). Improved estimators for a general class of beta regression
models. {\it Computational Statistics \&  Data Analysis}, {\bf 54}, 348--366. 

\R{Smithson, M. \& Verkuilen, J.}~(2006). 
A better lemon squeezer? Maximum likelihood regression with beta distributed dependent variables.
{\it Psychological Methods}, {\bf 11}, 54--71. 

\R{Stasinopoulos, D. M. \& Rigby, R. A.}~(2007). Generalized additive models for location scale and shape (GAMLSS) in {\tt R}. {\it Journal of Statistical Software}, {\bf 23}, 1--43.


\R{Venables,  W. N. \& Ripley, B. D.}~(2002). {\it Modern Applied Statistics with {\tt S},} 4th ed.  New York: Springer.

\R{Wei, B. C.}~(1998). {\it Exponential Family Nonlinear Models}. Singapore: Springer.


\R{Yoo, S.}~(2004). A note on an approximation of the mobile communications expenditures distribution
function using a mixture model. {\it Journal of Applied Statistics,} {\bf 31}, 747--752.

\end{description}

\end{document}